\documentclass[aps,rmp,reprint,amsmath,amssymb,longbibliography,superscriptaddress]{revtex4-2}
\usepackage{graphicx}
\usepackage{hyperref}
\usepackage{bm}
\usepackage{physics}
\usepackage{dsfont}
\usepackage{mathrsfs}
\usepackage{tabularx}
\newcommand{\eref}{Eq.~\eqref}

\begin{document}

\title{Simulating Non-Markovian Dynamics in Open Quantum Systems}

\author{Meng Xu}
\email{meng.xu@uni-ulm.de}
\affiliation{
Institute  for Complex Quantum Systems and IQST, Ulm University, Albert-Einstein-Allee 11, 89069  Ulm, Germany}

\author{Vasilii Vadimov}
\email{vasilii.1.vadimov@aalto.fi}
\affiliation{
QCD Labs, QTF Centre of Excellence, Department of Applied Physics, Aalto University, P.O. Box 15100, 00076 Aalto, Finland}

\author{J. T. Stockburger}
\email{juergen.stockburger@uni-ulm.de}
\affiliation{
Institute  for Complex Quantum Systems and IQST, Ulm University, Albert-Einstein-Allee 11, 89069  Ulm, Germany}

\author{J. Ankerhold}
\email{joachim.ankerhold@uni-ulm.de}
\affiliation{
Institute  for Complex Quantum Systems and IQST, Ulm University, Albert-Einstein-Allee 11, 89069  Ulm, Germany}

\date{\today}

\begin{abstract}
Recent advances in quantum technologies and related experiments have created a need for highly accurate, versatile, and computationally efficient simulation techniques for the dynamics of open quantum systems. Long-lived correlation effects (non-Markovianity), system-environment hybridization, and the necessity for accuracy beyond the Born-Markov approximation form particular challenges. Approaches to meet these challenges have been introduced, originating from different fields, such as hierarchical equations of motion, Lindblad-pseudomode formulas, chain-mapping approaches, quantum Brownian motion master equations, stochastic unravelings, and refined quantum master equations. This diversity, while indicative of the field's relevance, has inadvertently led to a fragmentation that hinders cohesive advances and their effective cross-community application to current problems for complex systems. How are different approaches related to each other? What are their strengths and limitations? Here we give a systematic overview and concise discussion addressing these questions. We make use of a unified framework which very conveniently allows to link different schemes and, this way, may also catalyze further progress. In line with the state of the art, this framework is formulated not in a fully reduced space of the system but in an extended state space which in a minimal fashion includes effective reservoir modes. This in turn offers a comprehensive understanding of existing methods, elucidating their physical interpretations, interconnections, and applicability. 
\end{abstract}

\maketitle

\tableofcontents

\section{Introduction}
\label{sec:introduction}

\begin{figure}
 \centering
\includegraphics[width=8.6cm]{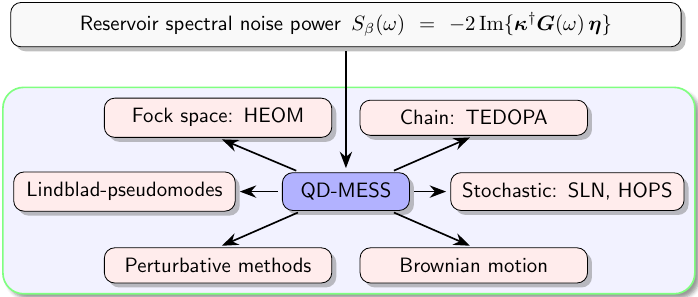}
\caption{Overview of several classes of embedding approaches for simulating non-Markovian dynamics in open quantum systems discussed in this Colloquium. As a general platform serves the formulation of time-local equations for quantum dissipation in minimally extended state space (QD-MESS). Its central ingredient is the spectral noise power $S_\beta(\omega)$ of a thermal reservoir and its decomposition in terms of Green's functions (top). The invariance of this representation with respect to linear transformations in reservoir mode space gives rise to a number of equivalent dynamical formulations which emerge from the QD-MESS (see arrows). These in turn give also rise to approximate treatments (bottom layer).}
\label{fig:qdmess-network} 
\vspace{-0.2cm}
\end{figure}

Realistic quantum systems inevitably interact with their environment, giving rise to open quantum system dynamics. These interactions induce irreversible processes such as decoherence, energy dissipation, transport in non-equilibrium stationary states, and quantum phase transitions that permeate diverse fields ranging from atomic physics and quantum optics to condensed matter, quantum information, chemical physics, and biology. Over the past decade, this long-standing area of research (see e.g.\ \textcite{breuer02,gardiner2004quantum,weiss12}) has experienced renewed attention, driven by rapid advances in the manipulation of quantum systems for technological applications.

Designing next-generation quantum computing platforms and advanced quantum sensing protocols, for example, requires precise and robust simulation methods that transcend the limitations of perturbative approaches, while remaining manageable on conventional hardware without excessive resource requirements. When applying open-system theory to complex systems or systems with time-dependent driving, the preconditions taken for granted in the conventional master equation approach are easily violated, in particular, 
the validity of the Born-Markov approximation~\cite{breuer02,nathan2020universal,mozgunov2020completely,davidovic2020completely}. 
Even in a system which has \emph{prima facie} a weak environmental coupling, these preconditions can be violated in the case of very low temperature or structured environments (e.g., bandgap materials). 

On larger quantum systems, the investigation of correlated environmental effects  may require the study of system Hamiltonians with dense energy level structure and thus with transition frequencies too small for the conventional secular approximation to apply. Incorporating time-dependent control fields into the standard master equation approach with full generality, e.g.\ for dynamical decoupling \cite{viola99} in quantum computing devices ~\cite{west10,sar12,liu2013noise},  poses another considerable challenge. Even without driving, the Lindblad operators constituting the dissipative terms of a conventional master equation are not system-agnostic as they depend on its eigenstates and level structure. With driving, this dependence on system properties becomes non-local in time~\cite{kallu22,wu22}. 

Some progress has been achieved within the Born–Markov regime in formulating Lindblad descriptions without invoking the secular approximation \cite{kirsanskas2018phenomenological,nathan2019topological,nathan2020universal,mozgunov2020completely,mccauley2020,davidovic2020completely,davidovic2022geometric,dabbruzzo2023a,shiraishi2025quantum}. Examples include the Universal Lindblad Equation~\cite{nathan2020universal}, the coarse-grained master equation~\cite{mozgunov2020completely}, and the geometric-arithmetic master equation~\cite{davidovic2020completely}. While these treatments remain valid irrespective of the system's many-body level structure and a time-dependent Hamiltonian \cite{davidovic2020completely,nathan2020universal,garcia2024fate,chen2022hoqst,aghaee2025interferometric,shiraishi2025quantum}, they are only applicable as long as the reduced dynamics can be considered to be Markovian but fail, for example, for strong system-bath coupling, at low-temperatures, for structured reservoirs, 
and for open many-body dynamics with time-retardation \cite{xu2024environment,min2024bath,windt2025effects,trivedi2024a}.

The common notion of a Markovian open-system dynamics thus becomes an unwieldy tool, at least when all parameter regions relevant to quantum phenomena and technologies are to be visited. This holds in particular when the formal description is to reflect an appreciable degree of detail of the physical characteristics of an environment, such as spectral features and forms of system-reservoir coupling.

There is one class of approaches to open quantum systems which avoids the construction of system-dependent dissipators altogether and, formally, provides an exact solution to the problem. 
Path integrals are an elegant method which works in this spirit, however, at the cost of introducing an action functional that is non-local in time~\cite{feynman63,weiss12}.
This type of path integral cannot be directly translated back into the language of canonical quantization and familiar equations of motion. Its numerical evaluation by direct sampling (Path Integral Monte Carlo simulations) is plagued by oscillatory integrands, the so-called dynamical sign problem \cite{suzuki1993quantum,egger94,egger2000path,muhlbacher2005nonequilibrium}, prompting the development of alternative schemes \cite{makri1995tensor,gull11,strathearn2018efficient,fux2021efficient,pollock2018non,bose2022multisite,cohen2015taming,cygorek2022simulation}. 

The fundamental obstacle of memory effects in any exact, reduced description has alternatively been addressed by various ``un-reduced'' descriptions. Formally, they restore time locality to the dynamics by \emph{embedding the system in a suitably enlarged state space} spanned by auxiliary degrees of freedom. These additional degrees of freedom thus introduced must, as a minimum, reproduce the reduced dynamics after they are eliminated. The degree to which these actually resemble a bona fide quantum system varies from approach to approach. Reducing these requirements to a bare minimum actually shows deep connections between existing disjoint efforts and reveals a unified perspective, as will be demonstrated below.

In frameworks of this type, a finite number of auxiliary degrees of freedom (with arbitrary spectral properties) is introduced to capture the reservoir’s time-retardation effects, and the resulting embedded systems can be treated either as closed or governed by effective Markovian dynamics. Historically, a variety of extended schemes have been proposed. For example, \textcite{mori1965,zwanzig1960ensemble,zwanzig1961memory,zwanzig2001nonequilibrium,peter2010markovian,kupferman2004fractional,andrew2013numerical} introduced auxiliary variables to recast generalized Langevin equations into memoryless forms, while \textcite{garg1985effect,marchesoni1982generalized,marchesoni1983extension,zusman1980outer} utilized auxiliary variables to obtain time-local Fokker-Planck descriptions. Other approaches include embedding the system within a Lindbladian framework by introducing pseudomodes, as discussed by \textcite{imamoglu1994,garraway1996cavity,garraway2006theory,dalton2001theory,dalton2003non-markovian,arrigoni2013nonequilibrium}. In addition, \textcite{tanimura89,tanimura1990nonperturbative,tanimura06} introduced effective modes to minimize bath retardation effects, resulting in an effective Liouville equation for the embedded system. Finally, \textcite{chin2010exact,hughes2009effective,bulla2005numerical,vega2015how} implemented a chain-topology of interacting modes, whereby the resulting system dynamics are described by time-dependent Schr\"odinger or Liouville equations. 

In the face of non-Markovian dynamics, 
recent advances demonstrate that embedding methods offer considerable advantages in terms of broad applicability, numerical stability, and computational efficiency. 
Moreover, rigorous error bounds for several embedding schemes, particularly regarding the treatment of environmental correlations, have recently been established~ \cite{huang2024unified,trivedi2021convergence,trivedi2022description,trivedi2024a,liu2025error}. 
These advantages have been further enhanced by substantial advances in computational power, machine learning, and many-body quantum techniques, including tensor network methods and quantum emulation.

In this Colloquium, we discuss several of the embedding approaches in current use, aiming to provide a concise overview and to illustrate how seemingly distinct formulations are intimately related, see Fig.~\ref{fig:qdmess-network}. Our focus is on environments with Gaussian fluctuations. This seems justified given the lack of experimental or theoretical knowledge about higher noise cumulants in most open quantum systems in fields such as condensed matter physics, quantum optics, atomic and molecular physics, quantum information processing, many-body physics, and quantum thermodynamics. This way, this Colloquium brings together approaches that have been developed in the last years, often independently of each other, in these fields given their specific backgrounds and actual realizations.

We introduce a comparably abstract umbrella theory for extended-state approaches, namely, quantum dissipation with minimally extended state space (QD-MESS), with a finite number of excitable harmonic degrees of freedom, taking a very general perspective on how \emph{statistical and spectral} features of the environment are to be translated into structural features of these fictitious modes. From this ``hub'' (Fig.~\ref{fig:qdmess-network}) different established approaches can be obtained by (i) canonical or Bogoliubov transformations for harmonics and (ii) unitary transformations in quantum state space. 

Our starting point leading to QD-MESS is the formally exact expression for the propagator of the reduced density operator in Liouville space, Sec.~\ref{Sec: OpenQuantumSystems}. By unraveling the corresponding time-nonlocal generator via a Gaussian ensemble average over effective modes, the propagator can be exactly mapped to a time-local evolution equation for a density operator in a minimally extended state space, Sec.~\ref{sec:qdmess}. This formulation serves as a platform for systematically connecting further alternative methods and clarifying their interrelations, as depicted in Fig.~\ref{fig:qdmess-network}. These methods include Hierarchical Equations of Motion (HEOM; Sec.~\ref{Sec:HEOM}), Lindbladian-pseudomode approaches (Sec.~\ref{sec:quasi-lindblad}), quantum Brownian motion master equations  (Sec.~\ref{Sec:wpsm}), stochastic approaches such as the Stochastic Liouville–von Neumann equation (Sec.~\ref{sec:sln}) and the Hierarchy of Pure States (Sec.~\ref{subsec:hops}). We then present the thermofield method as a particular case of recasting statistical properties into structural ones (Sec. \ref{Sec:thermofield}) before discussing the important subject of chain-mapping techniques (e.g., TEDOPA; Sec.~\ref{Sec:chain-topology}). While these approaches are, in principle, broadly applicable to diverse reservoir structures, temperatures, and coupling regimes, their computational efficiency can vary substantially. In Sec. (\ref{Sec:MasterEquation}) we establish a link between QD-MESS and the conventional master equation approach before offering a summary in Sec. (\ref{sec:summary}).

\section{Modeling of Open Quantum Systems}
\label{Sec: OpenQuantumSystems}

To set the stage, we briefly recall the common modeling of open quantum systems based on system plus bath models \cite{weiss12,leggett87,ford1988quantum}. The total Hamiltonian ($\hbar =k_\mathrm B = m = 1$) is given by
\begin{equation}\label{Eq:htot}
     \hat{H} =  \hat{H}_s +  \hat{H}_{\rm sb} +  \hat{H}_b \,,
\end{equation}
where a system of interest, whose intrinsic properties are described by $ \hat{H}_s$, interacts with a reservoir, typically a heat bath with bulk properties, characterized by Hamiltonian $\hat{H}_b$, governing free fluctuations, and a linear separable system-bath interaction 
\begin{equation}\label{Eq:Hsb}
 \hat{H}_{\rm sb} = \hat{S}\otimes \hat{X}_b \,,
\end{equation}
where $\hat{S}$ and $\hat{X}_b$ denote operators acting on the Hilbert state space of the system and bath, respectively. More complicated interactions can be expressed (or well approximated) as a sum over separable terms \cite{vega17}.

Considering an initially factorizing state $\hat{\rho}_{\rm sb}(0) = \hat{\rho}_s(0)\otimes \hat{\rho}_{b}$ with initial states of system $\hat{\rho}_s(0)$ and bath $\hat{\rho}_{b}$, respectively, the time-ordered and reservoir-averaged propagator of the system density matrix in the interaction representation is generally given by
\begin{align}\label{Eq:syspropgen}
\mathcal{J}(t) =\left\langle \mathcal{T} \exp\Bigl(-i\int_0^t d\tau \mathcal{L}_{\rm sb}(\tau)\Bigr) \right\rangle_{b} \, .
\end{align}
Here $\mathcal{L}_{\rm sb} \cdot = [\hat{H}_{\rm sb},\cdot]$ is the interaction Liouvillian superoperator and $\mathcal{T}$ denotes the usual Dyson time-ordering operator. The angle bracket indicates the projection to superoperators in the system Liouville space via the partial trace over the bath $\langle \cdot \rangle_{b}
= \Tr_{\rm b}\bigl\{ \cdot\,\hat\rho_{b}\bigr\}$. 

For reservoirs with macroscopically many degrees of freedom (bulk properties), the paradigm of a Gaussian environment always applies,
unless the dissipation is both intrinsically nonlinear and dominated by a small neighborhood of the system. Assuming an unbiased interaction, the propagating superoperator $\mathcal{J}(t)$ is thus rendered as~\cite{kubo62,feynman63,aurel20,huang2024unified}
\begin{equation}\label{Eq:sysproggauss}
\mathcal{J}(t) =  \mathcal{T} \exp{-\frac{1}{2}\!\int_0^t\!\! d\tau\!\! \int_0^t\!\! du \left\langle\mathcal{T}\mathcal{L}_{\rm sb}(\tau) \mathcal{L}_{\rm sb}(u) \right\rangle_{b} } .
\end{equation}
Applying $\mathcal{J}(t)$ to $\hat{\rho}_s(0)$ yields the reduced density matrix at a later time  
\begin{equation}\label{Eq:superrho}
    \hat{\rho}_s(t) = \mathcal{J}(t)\, \hat{\rho}_s(0) = \mathcal{T}\, e^{-i\Phi(t)}\, \hat{\rho}_s(0) \,.
\end{equation}
The time-ordered exponential represents the operator form of the Feynman-Vernon influence functional \cite{feynman63} with 
\begin{equation}\label{Eq:IFsuperop}
    \Phi(t) = \int_0^t\!\! d\tau\!\! \int_0^t\!\! du \begin{bmatrix}
        \mathcal{S}_{q}(\tau) & \mathcal{S}_{c}(\tau) 
    \end{bmatrix} 
\bm{\Sigma}(\tau,u)
    \begin{bmatrix}
        \mathcal{S}_{q}(u) \\ \mathcal{S}_{c}(u)
    \end{bmatrix} \, .
\end{equation}
Here,  the classical and quantum superoperators $\mathcal{S}_{c/q}$, respectively, are introduced
which act on the density operator according to $\mathcal{S}_{c/q}\hat{\rho}=(\hat{S}\hat{\rho}\pm\hat{\rho}\hat{S})/\sqrt{2}$.
The real time self-energy matrix 
\begin{equation}\label{Eq:self-energy_matrix}
    \bm{\Sigma}(\tau,u) = \begin{bmatrix}
        -i & -i\theta(\tau-u)  \\ i\theta(u-\tau) & 0
    \end{bmatrix} C(\tau,u)\,
\end{equation}
collects the contributions of the two-point correlation function $C(\tau,u)$. The latter describes the free fluctuations of the reservoir observable $\hat{X}_b$, which is no longer considered as a dynamical variable at this point. \eref{Eq:IFsuperop} contains the reservoir dynamics only implicitly in the form of a retarded self-interaction. The Heaviside step function $\theta(t)$ enforces causality: it equals $1$ for $t \ge 0$ and $0$ otherwise.

In the generic case of a bath in thermal equilibrium, the fluctuations are stationary and thus characterized by a single time argument,
\begin{equation}\label{Eq:cf_position}
C(t) = \mathrm{Tr}_{\rm b} \left\{\hat{X}_{b}(t)\hat{X}_{b}(0)\hat{\rho}_{b}\right\} \,.
\end{equation}
Since the thermal fluctuations follow the fluctuation-dissipation theorem \cite{Kubo1966the,weiss12}, the reservoir's spectral noise power
\begin{equation}\label{Eq:spectrum}
S_\beta(\omega)=\int_{-\infty}^{\infty} dt\, {e}^{i\omega t}\, C(t) \,
\end{equation}
can be obtained from the inverse reservoir temperature $\beta = 1/T$ and the dissipative response of the reservoir, i.e. 
\begin{equation}\label{Eq:spectralnoise}
S_\beta(\omega) = \frac{J(\omega)}{1- {e}^{-\beta\omega}} \, .
\end{equation}
Here, the spectral density $J(\omega)$ is introduced as an anti-symmetric function. In the time domain, the above relation leads to
\begin{align} \label{Eq:bath_correlation}
    C(t) = \frac{1}{2\pi}\int_{-\infty}^{\infty} d\omega\, S_\beta(\omega)\, {e}^{-i\omega t} \,.
\end{align}
We mention in passing that Eqs.~(\ref{Eq:spectrum})--(\ref{Eq:bath_correlation}) can formally accommodate stationary correlations beyond thermal equilibrium by allowing $\beta$ to be a function of $|\omega|$. 
With $C(-t)= C^*(t)$ and the symmetry of the integration domain in \eref{Eq:IFsuperop}, one obtains an alternative form~\cite{feynman63} of the self-energy matrix,
\begin{equation}\label{Eq:self-energy_matrixFV}
    \bm{\Sigma}_{\rm symm}(t) = -i \begin{bmatrix}
        \Re C(t) & i\theta(t) \Im C(t) \\ i\theta(-t) \Im C(-t) & 0
    \end{bmatrix},
\end{equation}
whose matrix elements are Keldysh, retarded, and advanced Green's functions of the `field' $\hat{X}_b$, respectively. There is yet another useful form of the self-energy used in the literature \cite{weiss12}, i.e., 
\begin{align} \label{Eq:self-energy_matrixRI}
\bm{\Sigma}'(t) &= -i \theta(t) \left(
C(t)
\begin{bmatrix}
    1 & 1\\ 0& 0
\end{bmatrix}
+ C^\ast(t)
\begin{bmatrix}
    1 & -1\\ 0& 0
\end{bmatrix}
\right) \\\nonumber
&= -2i \theta(t) \left(
\Re C(t)
\begin{bmatrix}
    1 & 0\\ 0& 0
\end{bmatrix}
+ \Im C(t)
\begin{bmatrix}
    0 & i\\ 0& 0
\end{bmatrix}
\right),
\end{align}
which renders the double integration in the definition \eref{Eq:IFsuperop} of the non-local phase $\Phi(t)$ time-ordered.

Historically, the reservoir model was often constructed based on elementary excitations such as microscopic bosonic modes, e.g., phonons or plasmons, or a phenomenological oscillator model. Then, the only information required about Gaussian thermal environments are the bath spectral density $J(\omega)$ and the temperature $T$. Knowledge about actual microscopic degrees of freedom is not necessary which implies wide applicability.

In essence, in the reduced framework, cf. Eqs.~\eqref{Eq:superrho} and \eqref{Eq:IFsuperop}, the self-energy $\bm{\Sigma}(\tau,u)$ induces self-interactions during the system's time evolution. 
Depending on temperature and spectral bath distribution, these can be very long-ranged which implies that in general a time-local evolution equation for the reduced density cannot be formulated. 
To illustrate this, let us consider an ohmic distribution of the form $J(\omega)\propto \omega$ up to a characteristic cut-off frequency $\omega_c$. In the classical limit, ohmic friction corresponds to memoryless viscous friction (Stokes friction). In the high temperature limit $\omega_c\beta\ll 1$, one estimates from \eref{Eq:spectralnoise} and \eref{Eq:bath_correlation} that $C(t)\propto \delta(t)$ (white noise with bandwidth $\approx \omega_c$) while at zero temperature $C(t)\propto 1/t^2$. The algebraic decay in the low temperature domain is a clear signature of the dominant role of low frequency quantum fluctuations in the reservoir. It is this long-range decay which turns precise simulations of open quantum system dynamics into a formidable challenge. In the frequency domain, long-time tails are related to the analytic structure of $S_\beta(\omega)$, understood as a complex function.
The poles in \eref{Eq:spectralnoise} (Matsubara poles) are regularly spaced on the imaginary axis with neighboring poles separated by $2\pi/\beta$. For vanishing temperature, $\beta\to\infty$, these poles merge into a branch cut for $T\to 0$, the frequency-domain signature of slower-than-exponential decay.

\subsection{Green's function representation of bath spectra}
\label{Sec:gfrbs}

The retardation effects described in the previous subsection are a major obstacle when simulation techniques beyond a direct numerical or Monte Carlo evaluation of the path integral. As in many branches of physics, retardation effects can be eliminated by formally introducing auxiliary degrees of freedom. Formally, the first step towards this goal consists of expressing the noise spectrum [\eref{Eq:spectrum}] using matrix-valued single-particle Green's functions $\bm{G}(\omega)$, i.e.,
\begin{equation}\label{Eq:spectrum_dcp}
\begin{split}
    S_{\beta}(\omega) & 
    =-2 \Im{\bm{\kappa}^\dagger \bm{G}(\omega) \bm{\eta}} \\
    &= i\bigl\{ \bm{\kappa}^\dagger \bm{G}(\omega)\,\bm{\eta} 
    - \bm{\eta}^\dagger \bm{G}^\dagger(\omega)\,\bm{\kappa} \bigr\} \,
\end{split}
\end{equation}
with constant coupling vectors $\bm{\kappa}$ and $\bm{\eta}$, and
\begin{equation}
    \bm{G}(\omega) = \left(\,\bm{\omega}\mathds{1} - \bm{\mathcal{E}} \,\right)^{-1} \,.
\end{equation}
Here the matrix \(\bm{\mathcal{E}}\) is not required to have any particular mathematical properties, except that its eigenvalues should have a non-positive imaginary part, reflecting the fact that correlation functions decay in the long-time limit. The matrix $\bm{\mathcal{E}}$ represents the spectral response structure of environmental correlations (mapped to the free decay of auxiliary modes). To fully capture the structure of \(S_{\beta}(\omega)\) in terms of Green's functions, $\bm{\mathcal{E}}$ is, in principle, of infinite size, which is obviously impractical. For example, a naive decomposition in terms of Matsubara frequencies tends for $T\to 0$ into a continuum of modes (branch cut; see Fig.~\ref{fig:poles}).

For practical simulations, a very successful strategy is to treat the right-hand side of \eref{Eq:spectrum_dcp} as a rational function that approximates \(S_\beta(\omega)\) within a prescribed global error tolerance \(\epsilon\) \cite{xu2022taming,nakatsukasa2018aaa,takahashi2024high,thoenniss2025efficient,valera2021aaa,conni2024aaa,zhang2025minimal}. Specifically, for any chosen \(\epsilon\), there exists 
a rank-\(K\) matrix \(\bm{\mathcal{E}} \in \mathbb{C}^{\scriptscriptstyle K \times K}\) along with the column vectors \(\bm{\kappa},\bm{\eta} \in \mathbb{C}^{K\times 1}\) such that
 the corresponding approximant \(S_\beta^{\scriptscriptstyle (K)}(\omega)\) satisfies the global bound \(|S_\beta^{\scriptscriptstyle(K)}(\omega)-S_\beta(\omega)|<\epsilon\).

This close match is performed on the real frequency axis, where small errors here translate into small errors of either $J(\omega)$ or $\beta$ at any given frequency. This is a controlled error of the same type an experimental study would incur due to parameter uncertainties. Off the real axis, the approximant may differ significantly from the analytic continuation of a noise power spectrum $S_\beta(\omega)$. This is of particular interest when $S_\beta(\omega)$ is provided as a closed expression. Then even the Matsubara pole structure implicit in \eref{Eq:spectralnoise} will be reshaped into a much simpler pole structure.
For an efficient simulation, the goal is to find a minimal $K$, i.e.\ a \textit{minimal number of effective harmonic modes} with parameters encoding the information on temperature and spectral bath properties. 

The time-domain correlation function follows from the proper Fourier transform of Eq.~\eqref{Eq:spectrum_dcp}
\begin{equation}\label{Eq:corr_dcp}
    C(t) = i\bigl[\bm{\kappa}^\dagger \bm{G}(t)\, \bm{\eta} - \bm{\eta}^\dagger \bm{G}^\dagger(-t)\, \bm{\kappa} \bigr] \,
\end{equation}
with  Green's function $\bm{G}(t) = -i\theta(t)\,e^{-i\bm{\mathcal{E}}t}$. Formally associating \emph{quantum} modes with $\bm{G}(t)$ plays a crucial role in unraveling the influence functional, as seen in Sec.~\ref{sec:qdmess} and Appendix~\ref{Sec:skpi}.

Creating a few-mode representation of the relevant aspects of a many-mode environment can be seen as a form of model order reduction, a general technique widely used in both classical and quantum dynamical systems \cite{schilders2008model,benner2017model,asztalos2024reduced,luchnikov2024controlling}. However, in our case, the starting point is not an explicit large-scale system of differential equations \cite{sidles2009practical,fan2024model,luchnikov2024controlling,grigoletto2025exact}, but the many-body dynamics implicit in the reservoir correlation function. In the context of open quantum systems,
model order reduction offers an efficient solution to identify a minimal set of effective harmonic reservoir modes together with the couplings to the system to accurately describe the features of a given noise spectrum Eq.~(\ref{Eq:spectrum_dcp}) over a given time scale. Although these modes often lack a direct physical meaning, the resulting time-local dynamical equations turn out to be computationally very efficient. 

\subsection{Structural and numerical aspects of effective modes}
\label{subsec:discretize-bath}

Evidently, the representation \eref{Eq:spectrum_dcp} is not unique, not even for fixed $K$. Any invertible linear transformation performed on the vectors $\bm{\kappa}$ and $\bm{\eta}$ leads to a set $\{\bm{\kappa'},\bm{\eta'},\bm{\mathcal{E}'}\}$ with
\begin{equation}
\bm{\kappa'}^\dagger \bm{G'}(\omega) \bm{\eta'}
= \bm{\kappa}^\dagger \bm{G}(\omega) \bm{\eta} \,.
\end{equation}
This freedom can be exercised in various ways with different goals, e.g., to make either the vectors $\bm\kappa$ and $\bm\eta$ or the matrix $\bm{\mathcal{E}}$ sparse. An obvious, more specific choice consists in diagonalizing $\bm{\mathcal{E}}$ or, more generally, transforming it into Jordan normal form. We will show in the sequel that some of these choices correspond to a number of known special cases of time-local open-system quantum dynamics with extended states. Differences arise from different physical motivation, apart from the abstraction \eref{Eq:spectrum_dcp}. Not fully exercising the freedom allowed by \eref{Eq:spectrum_dcp} may, however, result in an increased number of modes $K$, which is detrimental to computational performance.

We also note that from a practical point of view, for a specific choice of effective reservoir modes and their couplings relaxing certain physical constraints on the individual mode level (e.g., odd symmetry of $J(\omega)$ or strictly positive couplings) can be tolerated as long as the aggregate of modes preserves physical consistency. This allows one to capture the entire temperature-dependent noise spectrum with significantly fewer auxiliary modes. Even the zero-temperature limit  $\lim_{\beta\to\infty} S_{\beta}(\omega) = \theta(\omega)\, J(\omega)$ can thus be very accurately described with a finite number of modes $K$, see e.g.~\cite{xu2022taming}.

\begin{figure}
    \centering
    \includegraphics[width=8.6cm]{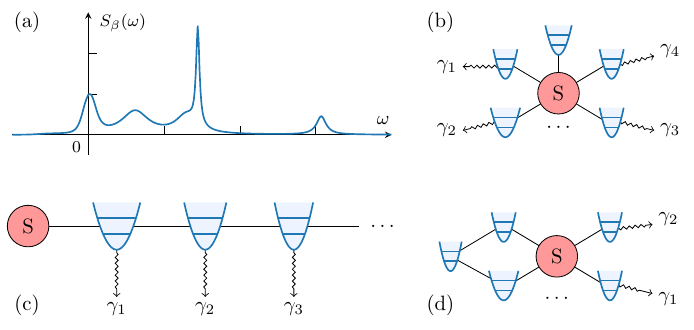}
    \caption{Diagram illustrating various system-bath configurations for efficient simulation schemes. A given spectral noise power $S_\beta(\omega)$ (a) can be decomposed into (b) a star topology (system interacts independently with each of the effective reservoir modes), into (c) a chain topology (system couples only to the first site of an interacting effective modes chain), or into (d) a mixed topology. Each topology is characterized by \(\bm{\mathcal{E}}\), $\bm{\kappa}$, and $\bm{\eta}$, see Eq.~\ref{Eq:spectrum_dcp}.}
\label{fig:mapping_model}
\end{figure}

With respect to coupling topology, several settings (see Fig.~\ref{fig:mapping_model}) have been considered in the literature with the most prominent ones corresponding to the extreme limits of the so-called star and the chain topology, respectively. The star topology refers to the situation where each of $K$ independent effective modes (diagonal $\bm{\mathcal{E}}$) couples directly to the system, while in the chain topology (tridiagonal $\bm{\mathcal{E}}$) the system-bath coupling only appears through a single mode which terminates a chain of $K$ interacting modes. We mention at this point that the general QD-MESS framework, which will be derived in the next Sec.~\ref{sec:qdmess}, does not refer to any topology, while certain representations that we discuss below do, for example, Sec.~\ref{Sec:chain-topology} is exclusively devoted to the chain topology.

{\em Star topology.---} This modeling is characterized by
a diagonal matrix $\bm{\mathcal{E}}$ with entries $\mathcal{E}_{jk} = \delta_{jk}(\omega_k - i\gamma_k)$, $\gamma_k \ge 0$, see Fig.~\ref{fig:mapping_model}(b), so that Green's function reduces to
\begin{equation}
    G_k(\omega) 
    = \frac{1}{\,\omega - \omega_k + i\,\gamma_k},\quad k = 1,2,\ldots,K\, .
    \label{eq:underdamped-GR}
\end{equation} 
To explicitly determine the mode parameters $\{\omega_k, \gamma_k\}$ and their couplings $\kappa_k$ and $\eta_k$, let us present   the so-called barycentric representation of rational functions which has the distinct merit of being numerically stable. It starts from a decomposition of the spectral noise power
\begin{equation}\label{Eq:rationbary}
 R_m(\omega) = \left. \sum_{j=1}^{m} 
 \frac{\theta_j S_{\beta}(\omega_j)}{\omega-\omega_j}   \middle/
 \sum_{j=1}^{m}\frac{\theta_j}{\omega - \omega_j} \right.
\end{equation}
with $R_m(\omega_j) = S_\beta(\omega_j)$. The key steps of the pole optimization algorithm are as follows ~\cite{nakatsukasa2018aaa,nakatsukasa2020algorithm}: (i) Setting a dense grid of candidate support points $\mathscr{J} = \{\omega_1,\omega_2,\ldots\}$ to represent the real-frequency axis. (ii) Starting from $m=1$ and at each iteration $m$, select the next support point $\omega_m$ as the maximizer of the current approximation error over the remaining grid $\mathscr{J}^{(m-1)} = \mathscr{J}\setminus\{\omega_1,\dots,\omega_{m-1}\}$, i.e., $\omega_m = \arg\max_{\omega\in \mathscr{J}^{(m-1)}} \bigl|S_\beta(\omega) - R_{m-1}(\omega)\bigr|$. (iii) Updating the approximant $R_m(\omega)$ via \eref{Eq:rationbary}, and compute the weights $\theta_j$ by minimizing the mean-square error of $R_m(\omega)$ over the set $\mathscr{J}^{(m)}$. The iteration terminates when the largest error is below a preset tolerance.

\begin{figure}
\centering
\includegraphics[width=8.6cm]{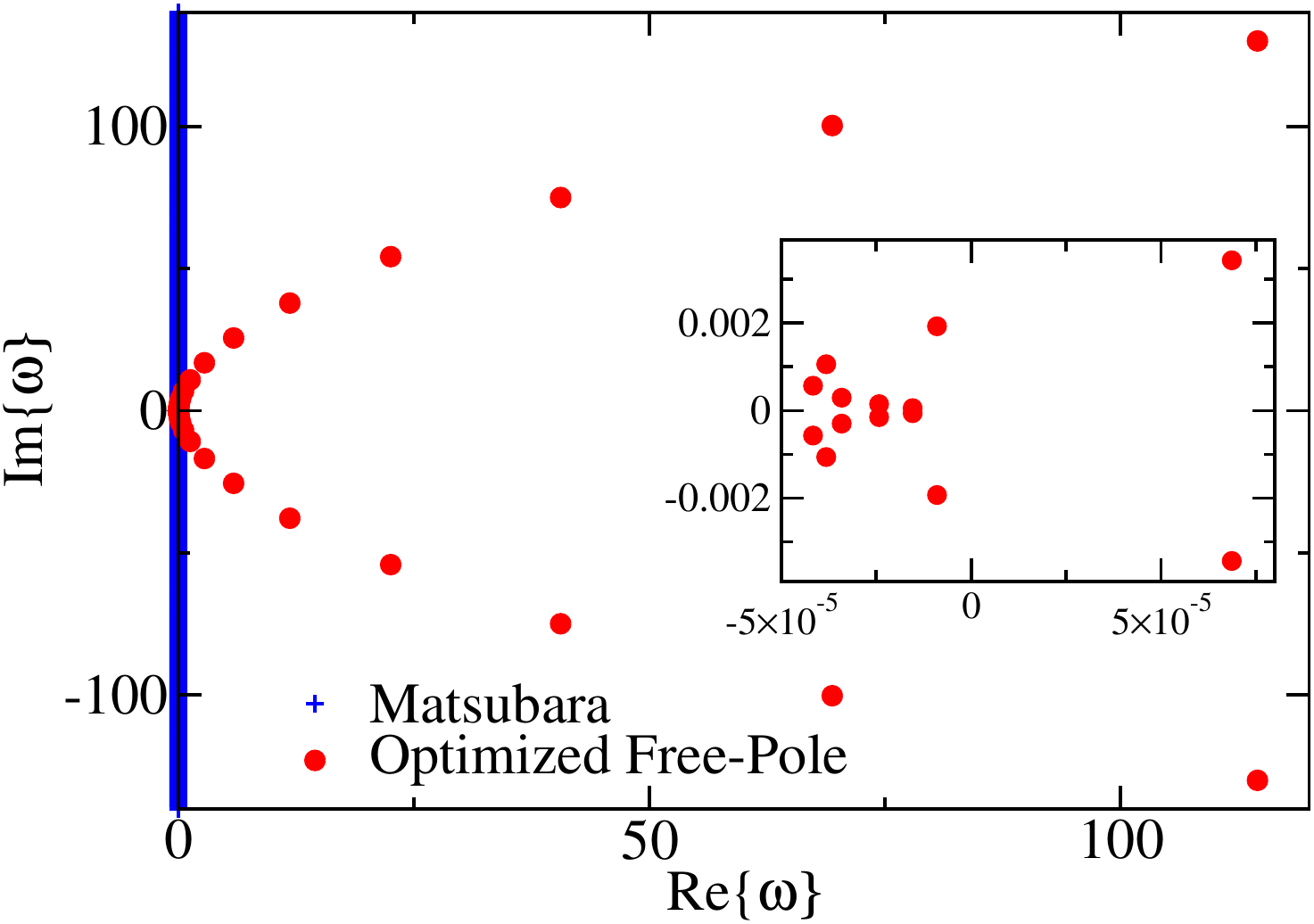}
\caption{Matsubara poles of the spectral noise power $S_\beta(\omega)$ of a thermal reservoir at $T=0$ that merge into a branch cut along the imaginary axis (blue thick vertical line) together with the optimized pole distribution (red dots) based on its rational decomposition [see Eq. (\ref{Eq:rationbary})]. Data refer to a subohmic spectral density $J(\omega)\propto \alpha \omega^s$ with $s=1/2$, $\alpha=0.05$ (in arbitrary units). Inset: low frequency range. From \textcite{xu2022taming}.}
\label{fig:poles}
\end{figure}

Even at $T=0$, the number of poles at which the procedure terminates grows rather slowly with increasing demand for accuracy, see \textcite{xu2022taming} for details. With pole locations and residues determined by the algorithm, a multi-exponential form $(t\geq 0)$ of the bath correlation function
\begin{equation}
\label{Eq:cbary}
    C(t) = \sum_{k=1}^{K} \,d_k\,  {e}^{-i\,(\omega_k -i\gamma_k)\, t}+\delta C(t)
\end{equation}
is recovered from the poles of $R_m(\omega)$ located in the lower complex half plane with correction $\delta C$ having negligible impact over a given time scale \cite{huang2024unified}. An example is shown in Fig.~\ref{fig:poles} for a subohmic model $J(\omega)\propto \omega^{1/2}$ at $T=0$ with $K=31$ and high accuracy, for details see \textcite{xu2022taming}. The distribution of quasimode parameters reflects the crucial low-frequency portion of $J(\omega)$, while it becomes sparse at higher frequencies with stronger "damping" (larger $\gamma_k$) reflecting a finite cut-off frequency. Note that this decomposition is system-agnostic and does not refer to any system-specific properties. Alternatively, one may work in the time domain by treating the right-hand side of \eref{Eq:cbary} as an approximation for \(C(t)\) and employing advanced algorithms to determine the corresponding parameters \cite{chen2022universal,takahashi2024high,takahashi2024finite,duan17,hartmann2019exact,olbrich2010time,lambert2019modelling}. 

Limiting cases of the general form \eref{Eq:cbary} include 
representations with either $\gamma_k = 0$ or $\omega_k=0$. In the former, the spectral density as a sum over delta-functions at frequencies $\omega_k$ is regained which is the starting point for approaches such as the  numerical renormalization group [NRG; \textcite{bulla2005numerical}], the density matrix renormalization group [DMRG; \textcite{ren2022time,xu2023hierarchical}], and the multiconfiguration time-dependent Hartree [MCTDH; \textcite{wang2003multilayer,wang10from,lindoy2021time}] by choosing linear, logarithmic, and nonuniform frequency meshes \cite{vega2015how}. In the latter, the representation in form of Matsubara frequencies is recovered which, as mentioned above and illustrated in Fig.~\ref{fig:poles}, implies an exponentially growing set of modes at lower temperatures~\cite{tanimura89,tanimura1990nonperturbative,hu2010communication,hu2011pade,ozaki2007continued,lambert2019modelling}. 

{\em Chain topology.---}
Tensor network methods have demonstrated significant numerical advantages in managing  exponentially large Hilbert spaces that arise in large-scale quantum system simulations. These methods rely on efficient representations of quantum states with low entanglement entropy between subsystems, such as matrix product states (MPS) for one-dimensional quantum systems.
In configurations with star topology, the entanglement entropy may scale with the number of bath modes, particularly under strong system-bath coupling or far-from-equilibrium dynamics. This nonlocal entanglement may require the use of large bond dimensions in tensor network simulations \cite{lacroix2024connectivity,li2022fly,kohn2021efficient,nuomin2024efficient,schroder2019tensor}.

To address this challenge, a common strategy has been to transform the star topology to a one-dimensional chain with nearest-neighbor interactions \cite{bulla2005numerical,hughes2009effective,chin2010exact,woods2014mappings,huh2014linear,tamura2007nonadiabatic,cederbaum2005short,gindensperger2007hierarchy,nazir2018the,nuomin2024efficient,shenvi2008efficient}, where only the terminating mode interacts directly with the system, see Fig.~\ref{fig:mapping_model}(c). While this chain topology facilitates the application of efficient tensor network algorithms such as time-evolving block decimation [TEBD; \textcite{vidal2004efficient,jaschke2018one,tamascelli2015improved}] and the DMRG, it is less efficient in determining a set of $K$ effective modes such that $K$ does not grow substantially for long-times in presence of structured environments. Further details of simulation methods based on the chain topology will be discussed in detail below
Sec.~\ref{Sec:chain-topology}.

Complementing the two aforementioned topological limits, advanced hybrid topologies [see Fig.~\ref{fig:mapping_model}(d)] have been investigated \cite{erpenbeck2018extending,rahman2019chebyshev,cui2019highly,ritschel2014analytic,hartmann2017exact,liu14,nakamura2018,vega2015how,ye2017low,meier1999non,mascherpa2020,medina2021few,lednev2024lindblad,lentrodt2020ab,monica2021accurate,huh2014linear,tamura2007nonadiabatic,cederbaum2005short,gindensperger2007hierarchy,derevianko2022exact,cui2025effective}. In numerical simulations, the optimal choice of topology depends critically on the specific properties of the system and the reservoir as well as on the relevant time scales.

\section{Quantum Dissipation with Minimally Extended State Space}
\label{sec:qdmess}

Following the decomposition~\eqref{Eq:corr_dcp} of the bath correlator into a set of effective modes (defined through the spectral properties of $\bm{\mathcal{E}}$), our aim is now an extension of the state space including these modes. In this section, we derive two principal forms of a system-mode Liouvillian $\mathcal{L}_{\rm sm}$, restoring time locality to the dynamics, with an extended state space which is computationally manageable. Propagation in the extended state space leads to the same propagating superoperator as given in Eqs.~\eqref{Eq:syspropgen} and~\eqref{Eq:sysproggauss}.

By inserting \eref{Eq:corr_dcp} into the first line of \eref{Eq:self-energy_matrixRI} the propagation superoperator in \eref{Eq:superrho} can be expressed as $\mathcal{J}(t) = \mathcal{T} \exp[-i \Phi_+(t) - i \Phi_-(t)]$ with 
\begin{subequations}\label{Eq:generating-action}
\begin{align}
\label{Eq:ga1}
    \Phi_{\!+}(t) &=\! \sqrt{2}\! \int_0^t \!\! d\tau du
    \, S_q(\tau)\,\bm{\kappa}^\dagger \bm{G}(\tau \!-\!u)\,\bm{\eta}\,\mathcal{S}_+\!(u)\\
    \Phi_{\!-}(t) &=\! -\sqrt{2}\! \int_0^t \!\! d\tau du
    \, S_{q}(\tau)\,\bm{\eta}^\dagger \bm{G}^\dagger(\tau\!-\!u)\,\bm{\kappa}\,\mathcal{S}_{\!-}(u) \,,
\label{Eq:ga2}
\end{align}
\end{subequations}
and superoperators $\mathcal{S}_\pm=(\mathcal{S}_c \pm \mathcal{S}_q)/\sqrt{2}$. 

In order to further specify the mode space, we have to interpret $\bm{G}$ and its adjoint as a generalized correlation function of modes with Gaussian fluctuations. This is accomplished through a bosonic state space associated with the matrix $\bm{\mathcal{E}}$, defined through a general mode Hamiltonian
\begin{align}
\label{eq:heff}
\hat{H}_{\rm mode} = \hat{\bm{a}}^\dagger \bm{\mathcal{E}}\, \hat{\bm{a}}\, ,
\end{align}
where we introduce matrix notation 
\begin{align}
\hat{\bm{a}}^\dagger = 
\begin{bmatrix}
\hat{a}_1^\dagger &\ldots & \hat{a}_K^\dagger
\end{bmatrix}
\quad \text{and} \quad
\hat{\bm{a}} = 
\begin{bmatrix}
\hat{a}_1 \\[-4pt] \vdots \\[-4pt] \hat{a}_K
\end{bmatrix} \;.
\end{align}
The Heisenberg operators $\hat{\bm{a}}(\tau)$ and $\hat{\bm{a}}^\dagger(u)$ of the effective modes have the matrix $\bm{G}(\tau-u)$ as vaccuum-state correlator, i.e.
\begin{equation}    
    \bm{G}(\tau-u) = -i\left\langle \bm{0}|\mathcal{T}\, \hat{\bm{a}}(\tau) 
    \hat{\bm{a}}^\dagger(u)|\bm{0} \right\rangle\, 
\end{equation}
which naturally guarantees the causality incorporated in Green's function. 
Directly taking the Hermitian adjoint of this expression would reverse the time ordering. We work in the Liouville space, in which the forward ordering can formally be restored by elevating the raising and lowering operators to left and right superoperators $\hat{a}_{k,+}$ and $\hat{a}_{k,-}$, respectively, which, as Liouville-space superoperators, act as $\hat{a}_{k,+}\hat{\rho}=\hat{a}_{k}\hat{\rho}$ and $\hat{a}_{k,-}\hat{\rho}=\hat{\rho}\,\hat{a}_{k}$; analogously for the adjoints. By construction, this leads to
\begin{align}\label{Eq:gpm-def}
    {\bm{G}}(\tau-u) &= -i\Tr{ |\bm{0}\rangle\langle \bm{0}|
    \,
    \mathcal{T}\, \hat{\bm{a}}_{+}(\tau) \hat{\bm{a}}_{+}^\dagger(u)\,|\bm{0}\rangle\langle \bm{0}|} \notag \\
    &
    \equiv -i\left\langle  \mathcal{T}\, \hat{\bm{a}}_{+}(\tau) \hat{\bm{a}}_{+}^\dagger(u) \right\rangle_m 
    \,
\end{align}
as well as 
\begin{align}\label{Eq:gpm-conj}
    {\bm{G}}^\dagger(\tau-u) = i\left\langle  \mathcal{T}\, \hat{\bm{a}}_{-}(u) \hat{\bm{a}}_{-}^\dagger(\tau) \right\rangle_m \,.
\end{align} 
We note that in the above the time ordering refers to the individual elements of the matrix $\hat{\bm{a}}(\tau)\hat{\bm{a}}^\dagger(u)$. According to this construction, the {\em mixed correlators} between any left and right superoperators vanish. 

The averaging procedure given in \eref{Eq:gpm-def} (i.e., initial preparation $|\bm{0}\rangle\langle \bm{0}|$ and final projection onto $|\bm{0}\rangle\langle \bm{0}|$) defines Gaussian statistics for ordered operator products~\cite{kubo62}. This implies for any time-dependent test functions (source fields) $\bm{X}(\tau)$ and $\bm{Y}(\tau)$ the following identity
\begin{align}\label{Eq:genfunctlp}
&\Big\langle \mathcal{T} \exp\!\Big[\!-\!i\!\int_0^t\!\! d\tau\, \bm{X}^\dagger(\tau)\hat{\bm{a}}_+(\tau)
+\hat{\bm{a}}_+^\dagger(\tau) \bm{Y}(\tau)\Big] \Big\rangle_m  \qquad \notag \\
&= \exp\!\Big[\!-\!i\!\int_0^t\!\! d\tau du\, \bm{X}^\dagger(\tau)\, \bm{G}(\tau-u)\, \bm{Y}(u) \Big] \, .
\end{align}
Now, this identity is used to unravel $e^{-i\Phi_+(t)}$ in terms of mode operators $\hat{\bm{a}}_+(\tau)$ and $\hat{\bm{a}}_+^\dagger(\tau)$; an analogous procedure applies to $e^{-i\Phi_-(t)}$ using $\hat{\bm{a}}_-(\tau)$ and $\hat{\bm{a}}_-^\dagger(\tau)$.
Consequently, the propagating superoperator in Eq.~\eqref{Eq:syspropgen} can be equivalently recast as
\begin{equation}
    \mathcal{J}(t) = \left\langle \mathcal{T} \exp{-i\int_0^td\tau\, \mathcal{L}_{\rm sm}(\tau)} \right\rangle_m\,,
\end{equation}
with the system-mode Liouvillian
\begin{equation}\label{Eq:sm_liouville}
\begin{aligned}
\mathcal{L}_{\rm sm}(\tau) =&\, 
\mathcal{S}_q(\tau)\bm{\kappa}^\dagger\hat{\bm{a}}_+\!(\tau)
+ \sqrt{2} \mathcal{S}_+\!(\tau)\hat{\bm{a}}_+^\dagger(\tau)\bm{\eta}\\
& + \mathcal{S}_q(\tau)\hat{\bm{a}}_-^\dagger\!(\tau)\bm{\kappa}
- \sqrt{2} \mathcal{S}_-\!(\tau)\bm{\eta}^\dagger\hat{\bm{a}}_-\!(\tau)\, ,
\end{aligned}
\end{equation}
where mode operators couple in a linear fashion to system operators. 

Moving back to the Schr\"odinger picture leads to an equivalent time evolution equation in extended state space starting from an initial state $\hat{\rho}_{\rm sm}(0) = \hat{\rho}_{s}(0)\otimes|\bm{0}\rangle\langle \bm{0}|$ given by the \emph{first major form of QD-MESS},
\begin{equation}\label{Eq:qdmess-density}
\boxed{
\begin{aligned}
    \dot{\hat{\rho}}(t) =& - i [\hat{H}_s, \hat{\rho}(t)] -i\bigl[\hat{H}_{\rm mode}, \hat{\rho}(t) \bigr]_{\star} \\ 
     - & i \mathcal{S}_{q} \left\{\bm{\kappa}^\dagger\hat{\bm{a}} + \hat{\bm{a}}^\dagger \bm{\eta}, \hat{\rho}(t) \right\}_{\star} -i \mathcal{S}_{c}\left[\hat{\bm{a}}^\dagger \bm{\eta} , \hat\rho(t) \right]_{\star}
    \, .
\end{aligned}
}
\end{equation}
Here we have dropped the subscript from the system-mode density operator for simplicity. Further, we employed the short-hand notation  $[\hat{O}, \hat\rho]_{\star} = \hat{O}\,\hat{\rho} - \hat{\rho}\,\hat{O}^\dagger$ and $\{\hat{O}, \hat{\rho}\}_{\star} = \hat{O}\,\hat{\rho} + \hat{\rho}\,\hat{O}$.
Following the above construction, the reduced density operator at time $t$ follows from the extended system-mode state via
\begin{equation}\label{Eq:rhos_mode}
\hat{\rho}_{s} = \langle \hat{\rho}_{\rm sm}\rangle_{\rm m} = \langle {\bm{0}}| \hat{\rho}_{\rm sm}|{\bm{0}}\rangle \,.
\end{equation}
As we will see in Sec.~\ref{Sec:HEOM}, this first form of QD-MESS is closely related to several variants of the HEOM method. An alternative derivation of \eref{Eq:qdmess-density} is provided in Appendix~\ref{Sec:skpi}. 

The state space in the first form of QD-MESS is a physically meaningful Liouville space, the dynamics governed by $\mathcal{L}_{\rm sm}$, however, has several features that differ significantly from more physical-motivated approaches where an environment is modeled by harmonic modes with a secondary environment: The damping of the modes considered so far is only due to the non-Hermitian nature of the mode Hamiltonian, and the system state is determined by \eref{Eq:rhos_mode}, not by a partial trace. In order to get a view of QD-MESS dynamics in which the modes appear more similar to open quantum systems themselves, it is convenient to perform a Bogoliubov transformation of the effective modes.

\emph{Bogoliubov transformation.---}
For that purpose, we introduce the similarity transformation described by the superoperator
\begin{equation}\label{Eq:bogoliubov-keldysh}
\mathcal{B}
= \exp\!\left\{-\hat{\bm{a}}_{-}^\dagger \hat{\bm{a}}_{+} \right\}
  \exp\!\left\{\frac{1}{\sqrt{2}}\big(\hat{\bm{a}}_+^\dagger\hat{\bm{a}}_+ + \hat{\bm{a}}_-^\dagger\hat{\bm{a}}_-\big)\right\}\,,
\end{equation}
where the non-vanishing commutators among left/right superoperators are $[\hat{a}_{k,+},\hat{a}_{l,+}^\dagger] = [\hat{a}_{k,-}^\dagger, \hat{a}_{l,-}] = \delta_{kl}$.  These operators transform according to 
\begin{equation}
\begin{aligned}
\mathcal{B}\,\hat{a}_{k,+}^{\dagger}\,\mathcal{B}^{-1}
&=\frac{1}{\sqrt{2}}\bigl(\hat{a}_{k,+}^{\dagger}-\hat{a}_{k,-}^{\dagger}\bigr), 
&\,
\mathcal{B}\,\hat{a}_{k,+}\,\mathcal{B}^{-1}
&=\sqrt{2}\, \hat{a}_{k,+},\\[2pt]
\mathcal{B}\,\hat{a}_{k,-}\,\mathcal{B}^{-1}
&=\frac{1}{\sqrt{2}}\bigl(\hat{a}_{k,-}-\hat{a}_{k,+}\bigr), 
&\,
\mathcal{B}\,\hat{a}_{k,-}^{\dagger}\,\mathcal{B}^{-1}
&=\sqrt{2}\, \hat{a}_{k,-}^{\dagger}.
\end{aligned} \notag
\end{equation}
Now, we apply this similarity transformation to \eref{Eq:qdmess-density} and will denote the transformed density operator $\mathcal{B}\,\hat{\rho}(t)$ again by $\hat{\rho}(t)$. Splitting the mode matrix into Hermitian and skew Hermitian parts, $\bm{\mathcal{E}} = \mathbf{\Omega} - i\mathbf{\Gamma}$, we obtain the \emph{second major form of QD-MESS},
\begin{equation}
\label{Eq:qdmess-sk}
\boxed{%
\begin{aligned}
\dot{\hat{\rho}}(t)
=\;& -i\bigl[\hat{H}_{s}
+\hat{S}\bigl(\hat{\boldsymbol{a}}^{\dagger}\boldsymbol{\kappa}_{+}
+\boldsymbol{\kappa}_{+}^{\dagger}\hat{\boldsymbol{a}}\bigr)
+\hat{\boldsymbol{a}}^{\dagger}\boldsymbol{\Omega}\,\hat{\boldsymbol{a}},\,
\hat{\rho}\bigr] \\
& + i\,\boldsymbol{\kappa}_{-}^{\dagger}\bigl(2\,\hat{\boldsymbol{a}}\,\hat{\rho}\,\hat{S}
-\{\hat{S}\hat{\boldsymbol{a}},\hat{\rho}\}\bigr) + \mathrm{H.c.} \\
& +2\,\hat{\boldsymbol{a}}\,\boldsymbol{\Gamma}^T \hat{\rho}\,\hat{\boldsymbol{a}}^{\dagger} - \bigl\{\hat{\boldsymbol{a}}^{\dagger}\boldsymbol{\Gamma}\,\hat{\boldsymbol{a}},\,\hat{\rho}\bigr\} \,.
\end{aligned}}
\end{equation}
Here, the commutators and anticommutators (without subscript $\star$) have their ordinary meanings, and $\bm{\kappa}_\pm=(\bm{\kappa}\pm\bm{\eta})/2$. The effective modes begin in vacuum, while the reduced density matrix of the system is obtained by tracing these out, $\hat{\rho}_s(t)=\mathrm{Tr}_{\rm m}\{\hat{\rho}(t)\}$. In the special case of vanishing $\bm{\kappa_-}$ and positive $\Gamma$, this equation has an obvious Lindblad structure; however, its validity is not limited to this case. Equation~(\ref{Eq:qdmess-sk}) will serve as the platform for further discussions of pseudomode--Lindblad formulations (Sec. \ref{sec:quasi-lindblad}) as well as wave--function methods in several subsequent sections.

We emphasize again that reservoir properties (including temperature) are encoded in the effective mode parameters, i.e. the entries of the $K\times K$ matrix $\bm{\mathcal{E}}$ which captures spectral response and interaction topology of effective modes, while the $K$-component column vectors $\bm{\kappa}$ and $\bm{\eta}$ determine the system-effective-modes couplings and  formally encode associated fluctuations. In numerical simulations, one often vectorizes $\hat{\rho}$ via the Choi–Jamiołkowski isomorphism and solves the resulting Schr\"odinger-like equation in a doubled Hilbert space, cf.~\textcite{ke2022hierarchical,zwolak2004mixed,cui2015variational,yuan2021solving,kshetrimayum2017simple,compaioli2024quantum}.

We refer to the set of equivalent non-perturbative time-local equations of motion Eqs.~\eqref{Eq:qdmess-density} and \eqref{Eq:qdmess-sk}, physically equivalent to the superoperator expression \eref{Eq:superrho}, as \emph{quantum dissipation with minimally extended state space} (QD-MESS). It serves as a universal framework which allows one to easily relate various other approaches for non-Markovian quantum dynamics. In particular, it is independent of any chosen topology for the system-reservoir interaction encoded in the matrices \(\bm{\mathcal{E}}\), $\bm{\kappa}$, and $\bm{\eta}$ and thus applies to both star and chain settings. We also emphasize that the way dissipation is treated in the QD-MESS is system-agnostic in contrast to the procedure for dissipators derived in weak-coupling Born-Markov master equations.

\section{Hierarchical Equations of Motion}
\label{Sec:HEOM}
The Hierarchical Equations of Motion [HEOM; \textcite{tanimura89,tanimura06,tanimura2020numerically,yan2014theory,yan2016dissipation,jin2008exact}] refers to a class of non-perturbative approaches for simulating open quantum systems where in the original derivation the Feynman-Vernon influence functional (see Appendix \ref{Sec:skpi}) is replaced by a nested hierarchy of differential equations for auxiliary density operators (ADOs). For computational purposes, only a finite set of ADOs can be considered. In the original formulation \cite{tanimura89,tanimura1990nonperturbative,tanimura06}, the number of ADOs grows rapidly with the hierarchy depth as well as the mode number \cite{ishizaki05,shi2009efficient}, thus posing severe computational challenges for low-temperature environments, at stronger system-bath coupling, and for long reservoir memory times. Several strategies have been developed to address these limitations. Among these are truncation schemes limiting the hierarchy depth \cite{ishizaki05,fay2022simple,kong2015dissipaton,hou2015improving,yan2004hierarchical,song2015time,xu05,humphries2024role}, white-noise residue ansatz methods \cite{yan2014theory,xu2007dynamics,ding2011optimized,ishizaki05}, and  on-the-fly dynamical filtering algorithms \cite{shi2009efficient} as well as efficient decomposition schemes, cf.~Sec.~\ref{subsec:discretize-bath}. 

HEOM approaches have turned into a standard tool to very successfully treat highly non-trivial problems for a broad class of  open quantum systems. Recently, propagating HEOM via tensor network representations \cite{shi2018efficient,borrelli2019density,borrelli2021expanding,yan2021efficient,ke2022hierarchical,ke2023tree,link2024open,chen2025tree,lindoy2025pyttn}, neural network approaches \cite{cao2024neural,ye2025simulating}, quantum simulation protocols \cite{li2024toward,dan2025simulating}, and the semiclassical propagation \cite{shi2025hierarchical}, has shown its promise in alleviating the curse of dimensionality even further. We mention in passing that historically numerous variants of HEOM have been proposed, beyond the three representative forms reviewed below.

\emph{Generalized HEOM.---}
A unified approach applicable to all types of reservoirs, the entire temperature range as well as strong coupling emerges when interpreting the generalized HEOM through the lens of the QD-MESS. We introduce the number basis of the effective modes according to $\hat{a}_k^\dagger\hat{a}_k |m_k\rangle = m_k |m_k\rangle$, $m_k \in \mathbb{N}$ and denote $|\bm{m}\rangle = \otimes_{k=1}^K|m_k\rangle$. 
This leads to the definition of \emph{auxiliary density operators}
\begin{align} \label{Eq:f-expansion}
    \hat{\rho}_{\bf m,n}(t) = \langle \bm{m}| \hat{\rho}(t)| \bm{n} \rangle \,.
\end{align}
We now use the decomposition (\ref{Eq:cbary}) of the bath correlator, implying star topology, and represent the first major form of QD-MESS, \eref{Eq:qdmess-density}, in the number basis of the effective modes as  generalized HEOM
\begin{align}\label{Eq:baryheom4}
&\dot{\hat{\rho}}_{\bf m,n}(t) = -i[\hat{H}_s, \hat{\rho}_{\bf m,n}] -\sum_{k=1}^K\Bigl\{( m_{k} z_{k} + n_{k} z_{k}^{*} ) \hat{\rho}_{{\bf m,n}} \notag \\
&\; -i\sqrt{(m_k + 1)d_k} \bigl[\hat{S},\hat{\rho}_{{\bf m}_k^{\!+},{\bf n}} \bigr] -i\sqrt{m_k d_{k}} \hat{S}\, \hat{\rho}_{{\bf m}_k^{\!-},{\bf n}} \notag \\[1mm]
&\;-i \sqrt{(n_k + 1) d_k^*}\bigl[\hat{S}, \hat{\rho}_{{\bf m,n}_k^{\!+}} \bigr] + i \sqrt{n_k d_{k}^{*}}\, \hat{\rho}_{{\bf m,n}_k^{\!-}}\hat{S} \Bigr\} \,.
\end{align}
This equation has been derived by various authors using different means \cite{suess2015hierarchical,link2022non,xu2022taming,vilkoviskiy2024bound,yan2004hierarchical,liang2024purified,zhang2025purified,su2025nonperturbative}. Its full potential as a computational tool, however, is only realized in combination with recent numerical optimization techniques yielding a small residual $\delta C(t)$ with a modest number of modes \cite{xu2022taming,chen2022universal,takahashi2024high}.

The coefficients $d_k$ and $z_k=\gamma_k + i\omega_k$ refer back to the multi-exponential decomposition \eqref{Eq:cbary}. The connection to our general expression \eqref{Eq:corr_dcp} is made through $\kappa_k^\ast =-\sqrt{2d_k}$ and $\eta_k = -\sqrt{d_k/2}$. The multi‐index $\bm m_k^\pm$ is obtained from $\bm m$ by raising (lowering) its $k$-th component by one. The ADOs \(\hat{\rho}_{\mathbf{m,n}}(t)\) retain operator status in the system’s Liouville space and the boundary conditions are inherited from the QD-MESS such that all ADOs with any $m_k,n_k \neq 0$ are initially set to zero, and the physical reduced density operator is \(\hat{\rho}_{s}(t) = \hat{\rho}_{\mathbf{0,0}}(t)\).

Alternatively, one can start from the bath correlator decomposition of \textcite{ikeda2020generalization}, 
\begin{equation} \label{Eq:ct_real_imag}
\begin{aligned}
    \Re\bigl\{C(t \ge 0)\bigr\} &= i\,\bm{\kappa}^\dagger \bm{G}(t)\,\bm{\eta}',\\[1mm]
    \Im\bigl\{C(t \ge 0)\bigr\} &= \bm{\kappa}^\dagger \bm{G}(t)\,\bm{\eta}''\,,
\end{aligned}
\end{equation}
which is related to \eref{Eq:corr_dcp} through $\bm{\eta} = \bm{\eta}' + \bm{\eta}''$.
The influence functional phase [see Eqs.~\eqref{Eq:IFsuperop} and \eqref{Eq:self-energy_matrixRI}] is thus recast as $\Phi(t) = \int_0^t d\tau du\, \bm{X}^\dagger(\tau)\, \bm{G}(\tau-u)\, \bm{Y}(u)$ with source fields $\bm{X}^\dagger = \sqrt{2} \mathcal{S}_q \bm{\kappa}^\dagger$ and $\bm{Y} = \sqrt{2}(\bm{\eta}' \mathcal{S}_q + \bm{\eta}'' \mathcal{S}_c)$. Green's function $\bm{G}(\tau-u) = -i\theta(\tau-u)\, e^{-i\bm{\mathcal{E}} (\tau-u)}$ is identified as the Hilbert-space time-ordered Green’s function of the effective modes. Following the unraveling of Sec.~\ref{sec:qdmess}, one obtains a time-local equation of motion
\begin{align}\label{Eq:qdmess-hilbert}
    \dot{\hat{\rho}}_w(t) 
    =&-i\,[\hat{H}_{s},\,\hat{\rho}_w(t)] -i\bigl[\hat{S},\,\bigl(\bm{\kappa}^\dagger\, \hat{\bm{a}} +\hat{\bm{a}}^\dagger\,\bm{\eta}'\bigr)\,\hat{\rho}_w(t)\bigr] \notag \\ &-\hat{\bm{a}}^\dagger\,\bm{\mathcal{E}}\,\hat{\bm{a}}\,\hat{\rho}_w(t)- i\bigl\{\hat{S},\,\hat{\bm{a}}^\dagger\,\bm{\eta}''\, \hat{\rho}_w(t)\bigr\} \,.
\end{align}
In the Fock representation of the effective modes, $\hat{\rho}_w(t) = \sum_{\bm{n}} \hat{\rho}_{\bm{n}}(t) |\bm{n}\rangle$, the projector follows
\begin{align} \label{Eq:ikeda-heom}
    \dot{\hat{\rho}}_{\mathbf{n}}(t) =& -i[\hat{H}_s, \hat{\rho}_s] - i\!\sum_{j,k=1}^K\!\! \sqrt{n_j (n_k + 1)}\, \mathcal{E}_{jk} \hat{\rho}_{\mathbf{n}_{jk}^{\!-+}} \notag \\[-2mm]
    & - i\sum_{j=1}^K \left[\hat{S}, \kappa_j^* \sqrt{n_j +1} \hat{\rho}_{\mathbf{n}_j^{\!+}} + \eta'_j \sqrt{n_j} \hat{\rho}_{\mathbf{n}_j^{\!-}}\right] \notag \\[-1mm]
    & - i \sum_{j=1}^K \left\{\hat{S}, \eta''_j \sqrt{n_j} \hat{\rho}_{\mathbf{n}_j^{\!-}}\right\} \,.
\end{align}
The multi-index $\bm{n}_{jk}^{-+}$ denotes the $j$-th element is lowered by one, while the $k$-th element is raised by one, relative to $\bm{n}$. 

This HEOM has a number of advantages: (i) It facilitates a transparent mapping to Brownian motion master equations (see Sec.~\ref{Sec:wpsm}); (ii) for certain spectral densities, it avoids instabilities that appear in the standard HEOM [\textcite{ikeda2020generalization}, see next subsection]; and (iii) in decompositions of bath correlation functions which involve positive- and negative-frequency components, i.e.\ \(\pm\,\omega - i\gamma\) \cite{ikeda2020generalization,li2022low,yan2014theory,yan2016dissipation,li2023dissipatons,chen2022universal,valera2021aaa,bradde2025modified}, this formulation requires only a single pair of creation--annihilation operators in contrast to \eref{Eq:baryheom4}, where each pole is treated independently. However, there is also the drawback that the decomposition may require a larger number of modes \cite{bradde2025modified}. As a result, the above HEOM may reduce computational costs in certain cases, but only if it is combined with a decomposition scheme which keeps the number of modes $K$ sufficiently small.

\emph{Standard HEOM.---} In the context of the star topology in Sec.~\ref{subsec:discretize-bath}, the original HEOM \cite{tanimura89,tanimura1990nonperturbative} immediately appears when setting all frequencies to zero ($\omega_k=0$) in the decomposition of \eref{Eq:cbary}. The generalized HEOM [Eqs.~\eqref{Eq:baryheom4} and \eqref{Eq:ikeda-heom}] then reduces to \cite{suess2015hierarchical,ikeda2020generalization,tanimura89} 
\begin{align}
\dot{\hat{\rho}}_{\bf n}(t) &= -i[\hat{H}_s,\hat{\rho}_{\bf n}] - i \sum_{k=1}^K\! \sqrt{\frac{n_k}{d_k}} \left( d_k \hat{S}\, \hat{\rho}_{{\bf n}_k^{\!-}} - d_k^\ast \hat\rho_{{\bf n}_k^{\!-}} \hat{S} \right) \notag \\
&- \sum_{k=1}^K n_k \gamma_k \hat{\rho}_{\bf n}  - i \sum_{k=1}^K \sqrt{(n_k + 1)d_k} [\hat{S}, \hat{\rho}_{{\bf n}_k^{\!+}}] \,.
\end{align}
Here, the ADOs have been rescaled \cite{shi2009efficient} such that higher-level ADOs decay to zero. This decay permits dynamical filtering based on the norm of the ADOs, thereby ensuring a preset error tolerance for the reduced density operator \(\hat{\rho}_s(t)\). Following our above discussion, the drawback of this equation is the growing number of effective modes required to achieve convergence beyond the domain of elevated temperatures and smooth spectral densities. 

As a final remark, we mention that recent studies \cite{dunn2019removing,ikeda2020generalization,krug2023stability,li2022low,yan2020new} have revealed that for specific spectral densities and parameters, the truncated HEOM propagation suffers from numerical instabilities due to emergent positive eigenvalues of the HEOM propagator. Mitigation strategies include modification of the ADOs basis \cite{yan2020new,li2022low,ikeda2020generalization} and reaction-coordinate mappings (chain topology) for effective modes \cite{ikeda2022collective}. 

\section{Lindblad-pseudomode Approach}
\label{sec:quasi-lindblad}

The Lindblad-pseudomode formalism has emerged as a powerful non-perturbative framework for open quantum systems by embedding discrete “pseudomodes” modeling a reservoir into a Lindblad-type master equation~\cite{imamoglu1994,garraway1996cavity,garraway1997nonperturbative,pleasance2020generalized,tamascelli2019efficient,medina2021few,monica2021accurate,luo2023quantum,lentrodt2023certifying}.  While this approach successfully captures non-Markovian dynamics, its current formulations rely on the restrictive assumption that the system-pseudomode couplings are {\em strictly positive} \cite{tamascelli2019efficient,lentrodt2023certifying}. This often necessitates a large number of pseudomodes \cite{huang2025coupled} which may cause computational inefficiencies, straining resources and limiting scalability for complex environments.

Extended frameworks propose to circumvent these limitations by employing effective modes with {\em complex-valued residues}~\cite{xu2022taming, chen2022universal, ritschel2014analytic, takahashi2024high, liu14, lambert2019modelling} , effectively projecting the parameter landscape onto a lower-dimensional manifold. This immediately raises the fundamental question if, and if yes how, such a generalized approach can be formulated to comply with a manifest Lindblad structure.

Transformations that rotate the system-bath interaction topology by introducing interacting pseudomodes
 have been explored to address this issue ~\cite{garraway1997nonperturbative, pleasance2020generalized, mascherpa2020,zhou2024systematic,lentrodt2023certifying}. In the rotated basis, the coupling parameters \(\bm{\kappa}\) and \(\bm{\eta}\) in \eref{Eq:spectrum_dcp} adopt positive values. Although analytical transformations remain tractable for systems coupled to only a few auxiliary modes~\cite{garraway1997nonperturbative, mascherpa2020, zhou2024systematic}, general cases involving multiple lossy modes pose significant challenges. Recently, interacting pseudomodes have also been employed in fitting procedures for bath spectra~\cite{lednev2024lindblad,medina2021few,monica2021accurate}.

Here, we consider the QD-MESS equation \eqref{Eq:qdmess-sk} which can be equivalently rewritten as 
\begin{align}
\label{Eq:qdmess-Lindblad}
    \dot{\hat{\rho}}(t) = - i[\hat{H}_0,\hat{\rho}] + \sum_{j,k=0}^{K} \mathcal{C}_{jk} \bigl(2\hat{f}_k \hat{\rho} \hat{f}_j^\dagger - \{\hat{f}_j^\dagger \hat{f}_k, \hat{\rho} \} \bigr)\,
\end{align}
with the self-adjoint system operator \(\hat{f}_0 = \hat{S}\) and effective mode operators \(\hat{f}_{k} = \hat{a}_k\) for $k\in \{1,2,3,\ldots, K\}$. The effective modes are initially in vacuum and the reduced density matrix of the system at time $t$ is obtained by tracing them out $\hat{\rho}_s(t)=\mathrm{Tr}_{\rm b}\{\hat{\rho}(t)\}$. The embedded system Hamiltonian 
\begin{align}\label{Eq:h-embedding}
  \hat{H}_0 = \hat{H}_s - \hat{S} (\hat{\bm{a}}^\dagger \bm{\kappa}_++ \bm{\kappa}_+^\dagger \hat{\bm{a}}) + \hat{\bm{a}}^\dagger \bm{\Omega}\, \hat{\bm{a}} \,
\end{align}
describes the system and effective modes interacting via matrices $\bm{\kappa}_+$ and $\bm{\Omega}$. The block Hermitian matrix 
\begin{align}
    \bm{\mathcal{C}} = \begin{bmatrix} 0 & i\bm{\kappa}_-^\dagger \\ -i\bm{\kappa}_- & \bm{\Gamma} \end{bmatrix}
\end{align}
governs the dissipation acting on the embedded system with parameter matrices related to the correlator decomposition Eq.~(\ref{Eq:spectrum_dcp}) via
\begin{align}
\bm{\kappa}_\pm = \frac{\bm{\kappa}\pm\bm{\eta}}{2},\quad
\bm{\Omega} = \frac{\bm{\mathcal{E}}+\bm{\mathcal{E}}^\dagger}{2},\quad
\bm{\Gamma} = \frac{\bm{\mathcal{E}}^\dagger-\bm{\mathcal{E}}}{2i}. 
\end{align}
When the parameters are identified by numerical fitting, one may enforce physically motivated structural constraints to null selected entries. Typical reductions include imposing $\bm{\kappa}_{-}=0$, which removes the off–diagonal dissipator blocks of $\bm{\mathcal{C}}$, and restricting $\bm{\Omega}$ and $\bm{\Gamma}$ to be diagonal. The latter case recovers the quasi–Lindblad pseudomode equation proposed by \textcite{park2024quasi}.  
 
One has to stress that \eref{Eq:qdmess-Lindblad} is \emph{not} of strict Lindblad form yet since \(\bm{\mathcal{C}}\) is not inherently semidefinite. A remedy involves to replace the $0$ in the upper left corner of  \(\bm{\mathcal{C}}\) by an effective damping rate \(\gamma_{s}\) \cite{thoenniss2025efficient}.

\emph{Lindblad-pseudomode formulation.---}
The Lindblad-pseudomode equation \cite{tamascelli2018nonperturbative,pleasance2020generalized,lentrodt2023certifying} is recovered from \eref{Eq:qdmess-Lindblad} by enforcing $\bm{\kappa}_-=0$, and a diagonal \(\bm{\mathcal{E}}\) . In particular, imposing in \eref{Eq:spectrum_dcp} the constraints $\bm{\kappa} = \bm{\eta}$ and $\mathcal{E}_{jk} = \delta_{jk}(\omega_k - i\gamma_k)$ yields 
\begin{align} \label{Eq:Lindblad-tamascelli}
    \dot{\hat{\rho}}(t) = -i[\hat{H}_0, \hat{\rho}] + \sum_{k = 1}^K \gamma_k \bigl(2\hat{a}_k \hat{\rho}\, \hat{a}_k^\dagger - \{\hat{a}_k^\dagger \hat{a}_k, \hat{\rho}\} \bigr) \,,
\end{align}
where $\hat{H}_0 = \hat{H}_s + \sum_k[ \hat{S} (\kappa_k \hat{a}_k^\dagger + \kappa_k^\ast \hat{a}_k) + \omega_k \hat{a}_k^\dagger \hat{a}_k]$ has a star topology system-pseudomode interaction, see Fig.~\ref{fig:mapping_model}(b). Allowing instead for non-diagonal entries, e.g.\ \(\mathcal{E}_{jk}=\omega_{jk}-i\,\delta_{jk}\gamma_k\), introduces interacting pseudomodes, see Fig.~\ref{fig:mapping_model}(c), and modifies the Hamiltonian to $\hat{H}_0 = \hat{H}_s + \sum_{jk}[ \hat{S} (\kappa_k \hat{a}_k^\dagger + \kappa_k^\ast \hat{a}_k)\delta_{jk} + \omega_{jk} \hat{a}_j^\dagger \hat{a}_k]$, as studied by \textcite{medina2021few,lednev2024lindblad,lentrodt2023certifying,mascherpa2020,zhou2024systematic,albarelli2024pseudomode}. 

In general, imposing constraints on the bath correlator decomposition to obtain non-interacting physical modes often results in a large number of effective modes $K$. By contrast, allowing for interacting modes may significantly reduce the mode number while retaining the simple Lindblad equation structure so that the dynamics can still be realized as a quantum channel \cite{medina2021few,lentrodt2023certifying,huang2025coupled}. Whether an equivalent interacting‑pseudomode representation exists for the general decomposition in \eref{Eq:spectrum_dcp}, remains an open question \cite{mascherpa2020,pleasance2020generalized,zhou2024systematic,huang2025coupled}.

\emph{Quasi-thermal decomposition.---} A second variant of the Lindblad–pseudomode approach encodes thermal fluctuations by casting the reservoir correlator in the form [see Eq.~\eqref{Eq:corr_dcp}]
\begin{equation}\label{Eq:diagonal-basis}
    C(t\ge0) = \sum_{k=1}^K g_k^2 \Bigl[(n_k + 1) {e}^{-i\omega_k t} + n_k {e}^{i\omega_k t}\Bigr] {e}^{-\gamma_k t} \,
\end{equation}
with \emph{fitting} parameters $g_k$, $n_k$, $\omega_k$, and $\gamma_k$ \cite{luo2023quantum,cirio2023pseudofermion,menczel2024nonhermitian,albarelli2024pseudomode,somoza2019dissipation}. Accordingly, one works within the finite temperature Green's function formalism (Keldysh formalism) so that the self-energy matrix in \eref{Eq:IFsuperop} takes the form
\begin{align}\label{Eq:standard_kmatrix}
    \bm{\Sigma}(t) = \sum_{j=1}^K g_j^2 \begin{bmatrix}
    G_j^K(t) & G_j^R(t) \\ G_j^A(t) & 0
    \end{bmatrix},
\end{align}
with the retarded $G_j^R(t) = -i\theta(t) e^{-i(\omega_j-i\gamma_j)t}$, the advanced $G_j^A(t) = i\theta(-t) e^{-i(\omega_j+i\gamma_j)t}$, and the Keldysh component $G_j^K(t) = (2n_j +1) [G_j^R(t) - G_j^A(t)]$. Analogous to the procedure in Sec.~\ref{sec:qdmess} and Appendix~\ref{Sec:skpi} for the unraveling procedure, one introduces modes with correlators related to the functions $G_j^K(t)$ and $G_j^{R/A}(t)$. However, instead of correlators of ``+'' modes, \eref{Eq:gpm-def}, or ``$-$'' operators, \eref{Eq:gpm-conj}, these functions are identified with correlations of ``c'' and ``q'' superoperators. They describe correlations of a thermal state, not the ground state. 
The notion of temperature is implied by interpreting $n_j$ as the occupation number of a bosonic mode with frequency $\omega_j$. 
 This way, one obtains an equation analogous to the second major form of QD-MESS, i.e.
\begin{align}\label{eq:lindblad-transformed}
    \dot{\hat{\rho}}(t) = -i[\hat{H}_0,\hat{\rho}] + &\sum_{k=1}^K\Bigl[ \gamma_k (n_k +1)\bigl(2\hat{b}_k\rho \hat{b}_k^\dagger - \{\hat{b}_k^\dagger \hat{b}_k,\hat{\rho}\}\bigr) \nonumber\\
    & + \gamma_k\,n_k\bigl(2\hat{b}_k^\dagger\hat{\rho} \hat{b}_k - \{\hat{b}_k \hat{b}_k^\dagger,\hat{\rho}\}\bigr)\Bigr] \,
\end{align}
with embedded Hamiltonian $\hat{H}_0 = \hat{H}_s + \sum_k [\omega_k\,\hat{b}_k^\dagger \hat{b}_k + g_k\,\hat{S}(\hat{b}_k+\hat{b}_k^\dagger)]$.
For an alternative derivation of \eref{eq:lindblad-transformed} see Appendix \ref{sec:thermofield-transformation}.

Note that in contrast to the proper thermofield approach using a physical thermal distribution (see Sec.~\ref{Sec:thermofield}), here, 
the reservoir is modeled by a number of $K$ modes initially occupied according to parameters $n_k$ determined by fitting \eref{Eq:diagonal-basis} to $C(t)$. Moreover, compared to \eref{Eq:Lindblad-tamascelli} where the pseudomodes are initialized  in vacuum (ground state), the above quasi-thermal initial state typically includes highly excited states which in turn degrades the simulation efficiency \cite{tamascelli2019efficient}. The two formulations are related by a so-called thermofield Bogoliubov transformation (see Appendix \ref{sec:thermofield-transformation}).

Specifically in quantum optics, where cavities with resonance frequencies $\omega_k$ and  photon leakage rates $\gamma_k$ exhibit long photon life-times, the spectral noise power can be modeled by narrow Lorentzians with $\omega_{k}\gg T, \gamma_k$ \cite{imamoglu1994,li2016transformation,delga2014quantum,varguet2019non,lentrodt2023certifying,gonzalez2014reversible,medina2021few}, i.e., 
\begin{equation} \label{Eq:SbetaLorentzian}
S_{\beta}(\omega) = \sum_k\frac{2\,\gamma_k g_k^2}{(\omega-\omega_k)^{2}+\gamma_k^{2}} \, .
\end{equation}
 In extended state space, this translates immediately into \eref{Eq:Lindblad-tamascelli}.

To efficiently propagate the above Lindblad schemes, tensor network methods \cite{somoza2019dissipation}, quantum simulations \cite{sun2025quantum,schlawin2021continuously,gajewski2025simulating}, and the multiple Davydov Ans\"atze \cite{zeng2025variational,yan2025multiple} have been developed. In a complementary scheme one can make use of pure-state unraveling methods such as the quantum jumps method \cite{xie2024photo}.

\section{Brownian Motion Master Equations}
\label{Sec:wpsm}

Phase-space representations recast the density operator as a quasi-probability distribution and map operator-valued  quantum dynamics into Fokker–Planck equations, providing a unified framework for quantum–classical correspondence. This formulation enables efficient semiclassical and mixed quantum–classical treatments \cite{martin2007semiclassical,stock1995semiclassical,shi2004semiclassical,koch2008non-markovian,provazza2019multi,wang1999semiclassical,montoya2017path,shi2025hierarchical}, and the resulting partial differential equations are amenable to classical algorithms and, more recently, physics-informed neural network solvers \cite{raissi2019physics,hu2025score,wang2025tensor}.

In principle, the QD-MESS in \eref{Eq:qdmess-hilbert} can be easily transformed into an equation of motion for phase-space variables via the squeezed representation
\begin{subequations}\label{eq:squeezed-ladders}
    \begin{align}
        \hat{a}_j &= \sqrt{2} \sigma_j q_j + \frac{1}{\sqrt{2}\sigma_j} \frac{\partial}{\partial q_j} \,, \\
        \hat{a}_j^\dagger &= -\frac{1}{\sqrt{2}\sigma_j} \frac{\partial}{\partial q_j} \,,
    \end{align}
\end{subequations}
where $q_j$ denotes a continuous variable corresponding to operator $\hat{q}_j = (\hat{a}_j + \hat{a}_j^\dagger)/\sqrt{2}\sigma_j$, and $\sigma_j$ is a nonzero scale factor to be detailed in the sequel. The right eigenfunctions of number operators $\hat{a}_j^\dagger \hat{a}_j$ in phase-space representation are 
\begin{align}\label{Eq:displaced_qbasis}
    \psi_{\bm{n}}^\beta(\bm{q}) &= \prod_{j=1}^K \pi^{-\frac{1}{4}} \sigma_j^{\frac{1}{2}} e^{-\frac{1}{2}\sigma_j^2 q_j^2 }\, \psi_{n_j}(q_j) \;\; \text{with} \\
    \psi_{n_j}(q_j) &= (\frac{\sigma_j}{\sqrt{\pi}\, 2^{n_j}\, n_j!})^\frac{1}{2}\, e^{-\frac{1}{2} \sigma_j^2 q_j^2}\, H_{n_j}(\sigma_j q_j) \notag \,,
\end{align}
where $H_n(\cdot)$ is the $n$th Hermite polynomial. Under the measure $d\mu(\mathbf{q})
  = \prod_j w_j^{-1}(q_j)\,dq_j$ where $ w_j(q_j)
  = \pi^{-\frac{1}{2}}\sigma_j\, e^{-\sigma_j^2q_j^2}$, these eigenfunctions form a complete and orthogonal basis  $\int_{-\infty}^{\infty} d\mu(\bm{q})\,  \psi_{\bf m}^\beta(\bm{q})\, \psi_{\bf n}^\beta(\bm{q}) = \delta_{\bf m,n}$. Accordingly, \eref{Eq:qdmess-hilbert} can be recast into a hybrid operator‐valued diffusion equation
\begin{align}\label{Eq:qdmess-fp}
    \dot{\hat{\rho}}_w =& -i[\hat{H}_s,\hat{\rho}_w] + i\sum_{j,k} \mathcal{E}_{jk}\frac{1}{\sigma_j}\frac{\partial}{\partial q_j} \bigl(\sigma_k q_k + \frac{1}{2\sigma_k}\frac{\partial}{\partial q_k} \bigr) \hat{\rho}_w \notag \\
    &- i\sum_j \bigl( \sqrt{2}\kappa_j^\ast \sigma_j q_j + \frac{\kappa_j^\ast - \eta'_j}{\sqrt{2}\sigma_j} \frac{\partial}{\partial q_j} \bigr) [\hat{S}, \hat{\rho}_w] \notag \\
    &+ i \sum_j \frac{\eta''_j}{\sqrt{2}\sigma_j}\frac{\partial}{\partial q_j} \{\hat{S}, \hat{\rho}_w\} \,,
\end{align}
with $\hat{\rho}_w \equiv \hat{\rho}_w(\bm{q},t)$ being an operator acting on the system Hilbert space and a $c$-number quasi‐distribution in effective‐mode space. The initial state of effective modes is set according to the distribution $\psi_{\bf 0}^\beta(\bm{q})$ and the reduced density operator is obtained by tracing out the full phase-space, i.e., $\hat{\rho}_s(t) = \int_{-\infty}^{\infty} d\bm{q}\, \hat{\rho}_w(\bm{q},t)$. The partial differential equation can be solved efficiently via classical finite‐difference schemes, offering an alternative to Fock‐state methods, especially when the effective modes become highly excited. In what follows, we focus on three specific Fokker-Planck equations arising from the decompositions [Eq.~\eqref{Eq:ct_real_imag}] that are particularly relevant to quantum optics, condensed‐matter physics, and chemical dynamics.

\emph{Extended Caldeira-Leggett Master Equation.---} 
We consider the  spectral density 
\begin{align}\label{Eq:brownianspectral}
    J(\omega) = \frac{2 c_0^2 \gamma_0\omega}{(\omega^2-\omega_0^2)^2+4 \gamma_0^2\omega^2}\, 
\end{align}
which, in the underdamped regime of $\omega_0 > \gamma_0$ and at high temperature $\beta \zeta \ll 1$ ($\zeta=\sqrt{\omega_0^2-\gamma_0^2}$), leads to
\begin{align}
    C(t\ge 0) = \frac{c_0^2}{\beta\omega_0^2} \phi_q(t) + i\frac{c_0^2}{2\omega_0} \phi_p(t) \, .
\end{align}
The functions $\phi_q(t) = {\rm e}^{-\gamma_0 t}[\cos{\zeta t} + (\gamma_0/\zeta)\sin{\zeta t}]$ and $\phi_p(t) = -(\omega_0/\zeta)~ {\rm e}^{-\gamma_0 t}  \sin{\zeta t}$ identify them as the position and momentum response of a damped oscillator~\cite[Sec.~7]{risken84}, respectively. Rewriting \(C(t)\) in the form of Eq.~\eqref{Eq:ct_real_imag} using nonvanishing components 
\[
\bm{\mathcal E}\!=\!
i\!\begin{bmatrix}
0 \!& \omega_{0}\\[-1pt]
-\omega_{0} \!& -2\gamma_{0}
\end{bmatrix},\;\;
\kappa_{1}=\eta_{1}'= \frac{c_0}{\omega_0}\beta^{-\frac{1}{2}},\;\;
\eta_{2}''=\frac{c_{0}}{2 i}\beta^{\frac{1}{2}},
\]
together with the scaling factors $\sigma_1 = \omega_0(\beta/2)^{\frac{1}{2}}$ and $\sigma_2 = (\beta/2)^{\frac{1}{2}}$, the general Eq.~\eqref{Eq:qdmess-fp} reduces to the Kramers-type Fokker-Planck equation \cite{caldeira1983path,garg1985effect} when identifying $\{q_1,q_2\} \rightarrow \{x,p\}$ , i.e.,
\begin{align}\label{eq:kramers_rho_w}
    \dot{\hat{\rho}}_w (x,& p,t) = -i[\hat{H}_s, \rho_w] - i c_0 \Big( [\hat{S}, x \hat{\rho}_w] + \frac{i}{2} \bigl\{ \hat{S}, \frac{\partial}{\partial p} \hat{\rho}_w \bigr\} \Big) \notag \\
    & + \Big( \omega_0^2 x \frac{\partial}{\partial p} - p \frac{\partial}{\partial x} + 2\gamma \frac{\partial}{\partial p} p + 2\gamma T \frac{\partial^2}{\partial p^2} \Big) \hat{\rho}_w \, .
\end{align}
This dynamical equation can also be put into  HEOM form \cite{li2022low,ding2017fokker,tanimura2015real} or into a Langevin equation \cite{peter2010markovian,kupferman2004fractional,marchesoni1983extension} which offers efficient simulation strategies for the above type of reservoir correlation. It is important to note that the specific form of the above  bath correlator decomposition limits the validity of the corresponding time evolution equations to sufficiently elevated temperatures. 

\emph{Extended Quantum Optical Master Equation.---}
In the strongly underdamped regimes, $\omega_0, \zeta \gg \gamma_0$, where the spectral distribution becomes sufficiently narrow, the Brownian spectral density in Eq.~\eqref{Eq:brownianspectral} reduces to
\begin{equation}
  J(\omega)\approx \frac{\gamma_0\,g^2}{(\omega-\zeta)^2+\gamma_0^2}\,, 
  \quad
  g^2=\frac{c_0^2}{2\,\zeta}\,.
  \label{eq:underdamped_J}
\end{equation}
This Lorentzian noise spectrum is commonly employed to model a single quantized mode of the electromagnetic field in a cavity \cite{imamoglu1994,li2016transformation,delga2014quantum,varguet2019non,lentrodt2023certifying,gonzalez2014reversible,medina2021few}, where \(\zeta\) denotes the cavity‐mode resonance frequency and \(\gamma_0\) its linewidth, corresponding to the photon‐leakage rate through the cavity mirrors. In the low‐temperature limit $\beta\zeta\gg 1$, one can approximate $S_{\beta}(\omega)\approx J(\omega)$ so that 
\begin{align}
    C(t \ge 0) &= g^2 e^{-i(\zeta - i\gamma_0) t} \,
\end{align}
can be recast in form of Eq.~\eqref{Eq:ct_real_imag} using a $2\times 2$ complex matrix $\bm{\mathcal{E}}$ with elements $\mathcal{E}_{11}= \mathcal{E}_{22} = -i\gamma_0$, $\mathcal{E}_{12}=-\mathcal{E}_{21}=i\zeta$, and non-vanishing coupling components $\kappa_1=\eta_1'= i\eta''_2= g$. Choosing the scaling factors $\sigma_1 = \sigma_2^{-1} = \zeta^{\frac{1}{2}}$, Eq.~\eqref{Eq:qdmess-fp} reduces to a Lindblad description \cite{isar1994open,vacchini2002quantum} of an extended system, when identifying $\{q_1,q_2\} \rightarrow \{x,p\}$ , i.e.,
\begin{align}
    \dot{\hat{\rho}}_w(x,p;t) =& -i[\hat{H}_s, \hat{\rho}_w] - ic_0 \bigl([\hat{S},x\hat{\rho}_w] + \frac{i}{2}\{\hat{S}, \partial_p \hat{\rho}_w \} \bigr) \notag \\
    & + \bigl( \zeta^2 x\partial_p -p\partial_x\bigr) \hat{\rho}_w + \gamma \bigl(\partial_x x + \partial_p p\bigr)\hat{\rho}_w \notag \\
    &+\frac{\gamma}{2}\bigl(\zeta\partial_p^2 + \frac{1}{\zeta}\partial_x^2\bigr) \hat{\rho}_w \,.
 \end{align}
Effectively, this equation describes a quantum subsystem coupled to a harmonic oscillator in the low temperature limit where the latter experiences friction and stochastic Gaussian noise through position and momentum. 

\emph{Smoluchowski-type equation.---}
The strong friction limit in classical physics, commonly referred to as the Smoluchowski limit \cite{smoluchowski1916brownsche,skinner1979derivation}, is characterized by a distinct separation of timescales: rapid equilibration in momentum compared to sluggish equilibration in position. This separation enables the adiabatic elimination of momentum from the Fokker–Planck equation, resulting in a time evolution equation for the marginal distribution in position only \cite{skinner1979derivation}.

In the limits of strong friction $\gamma_0 \gg \omega_0$ and sufficiently elevated temperature $\beta \ll 2\gamma_0/\omega_0^2$, the bath correlation function corresponding to Eq.~\eqref{Eq:brownianspectral} can be approximated on a coarse-grained timescale \(t \ge 2\gamma_0/\omega_0^2 \gg \beta, \gamma_0^{-1}\)  as
\begin{align}\label{Eq:smolucho_cf}
    C(t\ge 0) &= \left(\frac{c_0^2}{\beta\omega_0^2} - i\frac{c_0^2}{4\gamma_0} \right)\,  e^{-\frac{\omega_0^2}{2\gamma_0} t} \,,
\end{align}
which in turn recasts to Eq.~\eqref{Eq:ct_real_imag} using $\mathcal{E} = -i\omega_0^2/2\gamma_0$, together with $\kappa = \eta' = c_0 \beta^{-\frac{1}{2}}/\omega_0$, $\eta'' = -i\omega_0 c_0 \beta^{\frac{1}{2}}/4\gamma_0$. Choosing the scaling factor $\sigma = \omega_0\sqrt{\beta/2}$, Eq.~\eqref{Eq:qdmess-fp} reduces to a classical Smoluchowski equation \cite{garg1985effect,zusman1980outer,yang1989the,shi2009electron} 
\begin{align}
    \dot{\hat{\rho}}_{\rm S}(q,t) = & -i[\hat{H}_{s}, \hat{\rho}_{\rm S}] -ic_0[\hat{S}, q \hat{\rho}_{\rm S}] + \frac{c_0}{4\gamma_0} \big\{\hat{S}, \frac{\partial}{\partial q} \hat{\rho}_{\rm S}\big\} \notag \\
    &+ \frac{\omega_0^2}{2\gamma_0} \frac{\partial}{\partial q}(q + \frac{1}{\beta\omega_0^2} \frac{\partial}{\partial q}) \hat{\rho}_{\rm S} \, .
\end{align}
The generalization to the quantum regime \cite{ankerhold01,ankerhold2004low, maier2010quantum} is directly based on Eq.~\eqref{Eq:qdmess-fp}, e.g., keeping the leading order contribution from the Matsubara frequency. In this regime, the environment-induced linewidth significantly exceeds both the bare line separation and thermal excitations so that the dynamics of the harmonic mode exhibits nearly classical behavior but with substantial quantum fluctuations. 

For a two-level system that mimics, for example, donor–acceptor electron transfer in a Debye solvent at high temperatures, the above equation is equivalent to the Zusman equation \cite{zusman1980outer,garg1985effect,shi2009electron,yang1989the}. In the low-temperature regime, generalized Zusman equations have been derived \cite{ankerhold2004low,maier2010quantum,zhang2003theory,zhang2004low} that capture pronounced quantum fluctuations.

\section{Stochastic Liouville-von Neumann Equation}
\label{sec:sln}

In Sec.~\ref{sec:qdmess}, we recast the path integral representation of reduced density operator into deterministic time-local equations, i.e., QD-MESS in extended state space. We introduced auxiliary paths in order to replicate the effects of a many-particle environment by the time-local dynamics of effective reservoir modes closely resembling quantum-mechanical harmonic oscillators. In this section, we demonstrate that the effective model representing the environment can be \emph{classical}, although in the sense of classical probability, not classical physics.

This requires a pair of \emph{classical} Gaussian noise forces $Z_c(\tau)$ and $Z_q(\tau)$ such that propagation with the interaction Liouvillian 
\begin{equation}\label{Eq:stochLiou}
\mathcal{L}_{\rm noise}(\tau) =
Z_c(\tau)\mathcal{S}_q(\tau) + Z_q(\tau)\mathcal{S}_c(\tau)
\end{equation}
allows  the reduced density matrix $\hat\rho_{s}$ to emerge as a noise expectation value.
This is the case if these classical noise forces have autocorrelations and cross correlations such that
\begin{equation}\label{Eq:Zcorr}
\begin{bmatrix}
    \langle {Z}_c(\tau) {Z}_c(u) \rangle_{\rm av} & \langle {Z}_c(\tau){Z}_q(u)\rangle_{\rm av}\\
    \langle {Z}_q(\tau) {Z}_c(u)\rangle_{\rm av} & \langle{Z}_q(\tau) {Z}_q(u)\rangle_{\rm av}
\end{bmatrix}
= 2i{\bm{\Sigma}}_{\rm symm}(\tau-u),
\end{equation}
where $\langle \cdot \rangle_{\rm av}$ denotes the ensemble average over a classical Gaussian process and ${\bm{\Sigma}}_{\rm symm}$ is defined in \eref{Eq:self-energy_matrixFV}. At first glance, it seems impossible to find classical processes where $\langle{Z}_q(\tau) {Z}_q(u)\rangle_{\rm av}=0$ and $\langle {Z}_c(\tau){Z}_q(u)\rangle_{\rm av}\neq 0$. However, this difficulty can be overcome by allowing $Z_c(\tau)$ and $Z_q(\tau)$ to take complex values. Then $\langle{Z}_q(\tau) {Z}_q(u)\rangle_{\rm av}=0$ does not refer to an autocorrelation, since it does not contain any complex conjugate.

Using this generalization to complex-valued processes, we have a valid model of a \emph{quantum} environment with arbitrary $C(\tau)$ in terms of classical noise. 
Then open-system dynamics can be simulated by propagating the stochastic Liouville-von Neumann (SLN) equation~\cite{stockburger02}
\begin{align}\label{Eq:SLN}
    \dot{{\hat\rho}}_z(t) = -i[\hat{H}_s,\hat\rho_z] - i{Z}_{c}(t) \mathcal{S}_q {\hat\rho}_z - i{Z}_{q}(t)\mathcal{S}_c {\hat\rho}_z \,.
\end{align}

Path integrals form a mathematical language capable of describing both quantum probability amplitudes and classical probabilities; hence, path integral methods are often used as the basis of the SLN approach \cite{stockburger02,stockburger2004simulating,schmitz2019variance,mccaul2017partition}, where the basic concept is a simple Gaussian identity extended to functional integrals, without any reference to time ordering.

In practical applications, the sample average over $\sum_i \hat{\rho}_z^{(i)}/N$, performed after multiple independent propagations $\hat{\rho}_z^{(i)}$, provides an estimator for $\hat{\rho}_{s}$.
However, the following complication must be overcome for this approach to become workable: The Fourier transform of ${\bm{\Sigma}}_{\rm symm}$ does not yield matrices which are positive semidefinite, i.\,e.\ the r.h.s.\ of \eref{Eq:Zcorr} is not a classical correlation matrix.

On the other hand, when $Z_c(\tau)$ and $Z_q(\tau)$ are allowed to take complex values, \eref{Eq:Zcorr} can easily be fulfilled for any quantum environment with stationary Gaussian fluctuations. The noise properties \eqref{Eq:Zcorr} are not compatible with unitary propagation in \eref{Eq:SLN}. This prevents a direct physical interpretation of individual samples. Even for strong deviations from unitarity, the theory remains exact, but sampling based on it may become inefficient.

Since the complex conjugates of $Z_c(\tau)$ and $Z_q(\tau)$ do not enter \eref{Eq:Zcorr} at all, the statistical properties of $Z_c(\tau)$ and $Z_q(\tau)$ are underdetermined, allowing the freedom of different constructions of the noise pair. These are typically based on filtering white noise, one of them \cite{shao2004decoupling} can be linked to QD-MESS. We start from two complex white-noise processes $w_c(t)$, $w_q(t)$ whose only non-vanishing correlators are $\langle w_c^*(t) w_c(t')\rangle_{\rm av} = \langle w_q^*(t) w_q(t')\rangle_{\rm av} = \delta(t-t')$. Then the definitions\footnote{Strictly speaking, white noise should be treated using stochastic calculus. With the given correlations, and observing that no physical result depends on higher moments with respect to noise, the differences between ordinary calculus and its stochastic variants are immaterial.}
\begin{align}\label{Eq:filterShao}
Z_c(t) = w_c^*(t) &+ \int_0^t ds\, C(t-s) (w_c(s) + w_q(s))\nonumber\\
&+ \int_0^t ds\, C^*(t-s) (w_c(s) - w_q(s))\\
Z_q(t) = w_q^*(t)
\end{align}
satisfy \eref{Eq:Zcorr}. The mode decomposition \eref{Eq:corr_dcp} of the correlation function allows for a time-local description,
\begin{align}
Z_c(t) &= \bm{\kappa}^\dagger \bm{y}(t) + \bm{y}^\diamond(t) \bm{\kappa} + w_c^*(t)
\end{align}
with stationary solutions of the Ornstein-Uhlenbeck processes
\begin{align}
\partial_t{\bm{y}}(t) &= \bm{\eta}(w_c(t) + w_q(t)) - i\mathcal{E} \bm{y}(t)\\
\partial_t{\bm{y}}^\diamond(t) &= \bm{\eta}^\dagger (w_c(t) - w_q(t)) + i \bm{y}^\diamond \mathcal{E}^\dagger\,.
\end{align}

More systematic and computationally efficient constructions are obtained considering the $4\times4$
correlation matrix $\bm{\Sigma}_{\rm SLN}(\tau,\tau') = \langle \bm{{\tilde Z}}(\tau)\bm{{\tilde Z}}^\dagger(\tau')\rangle_{\rm av}$ with $\bm{{\tilde Z}} = [{Z}_c,{Z}_q,{Z}_c^\ast, {Z}_q^\ast]^T$.

This correlation matrix can be created by filtering real-valued white noise vectors $\bm{\xi}(t)$ with correlations $\langle\xi_m(t) \xi_n(t')\rangle_{\rm av} = \delta_{mn} \delta(t-t')$ using a $4\times 4$ filtering kernel,
\begin{equation}\label{Eq:noisefilter}
    \bm{\tilde Z}(t) = \int_{-\infty}^\infty \bm{\mathcal{F}}(t-t')\bm{\xi}(t').
\end{equation}
From the delta correlations, one concludes that
\begin{equation}
\bm{\Sigma}(t-t') = \int_{-\infty}^\infty dt'' \bm{\mathcal{F}}(t-t'') \bm{\mathcal{F}}^\dagger(t'-t'').
\end{equation}
This equation can be solved for $\bm{\mathcal{F}}$, the solution being non-unique. When choosing a causal filter, $\bm{\mathcal{F}}(\tau) =\theta(\tau) \bm{\mathcal{F}}(\tau)$ \cite{kailath2000}, the filter can again be decomposed into quasimodes in the spirit of QD-MESS. Then \eref{Eq:noisefilter} can be evaluated by propagating a system of simple linear stochastic differential equations.
Otherwise, the determination of $\bm{\mathcal{F}}$ and the convolution with white noise can be aided by applying the convolution theorem and implementing the filter with Fast Fourier Transform techniques.

It should be mentioned that stochastic simulations over long time intervals require significant numerical resources, which sometimes grow exponentially with the physical duration. This problem has been mitigated \cite{schmitz2019variance,stockburger2016exact,matos2020efficient,yan2022piecewise} but not fully solved for the general case.

\section{Hierarchy of Pure States}
\label{subsec:hops}

The Hierarchy of Pure States [HOPS;  \cite{suess14,suess2015hierarchical,hartmann2017exact,ritschel2015non}] has emerged as an efficient simulation method to solve the  Non-Markovian Quantum State Diffusion  equation (NMQSD) which has been developed a while ago \cite{diosi1998non,strunz1999open}. It is essentially a wave-function based method which is guided by the idea to introduce a set of auxiliary pure states to transform the functional derivative contained in the NMQSD and difficult to handle into a hierarchy of differential equations. Accordingly, the HOPS inherits the importance-sampling scheme from the NMQSD  \cite{suess14,hartmann2021open} and simultaneously employs a hierarchical structure analogous to that of the HEOM. In this way, HOPS leverages efficient bath correlation function decompositions and adaptive truncation schemes \cite{zhang2018flexible} to minimize the effective mode state space dimension.

The relation between QD-MESS and HOPS becomes transparent  when using the unnormalized number basis $|m)\equiv\sqrt{m!}\,|m\rangle$. Namely, for the bath correlator decomposition Eq.~\eqref{Eq:cbary}, one arrives for $\hat{\rho}_{\bf m,n}(t) = (\bm{m}| \hat{\rho}(t)|\bm{n})$ at the first major form of QD-MESS  
\begin{equation}
\label{Eq:baryheom5}
\begin{split}
&\dot{\hat{\rho}}_{\boldsymbol{m},\boldsymbol{n}}(t) = -i[\hat{H}_{s},\hat{\rho}_{\boldsymbol{m},\boldsymbol{n}}]
-\sum_{k=1}^{K}\Bigl[(m_k z_k + n_k z_k^{\ast})\,\hat{\rho}_{\boldsymbol{m},\boldsymbol{n}} \\
&-[\hat{S},\hat{\rho}_{\boldsymbol{m}_k^{\!+}\!,\boldsymbol{n}} \!-\!\hat{\rho}_{\boldsymbol{m},\boldsymbol{n}_k^{\!+}}]  
+ m_k d_k \hat{S}\,\hat{\rho}_{\boldsymbol{m}_k^{\!-}\!,\boldsymbol{n}}
+ n_k d_k^{\ast} \hat{\rho}_{\bm{m},\bm{n}_k^{\!-}} \hat{S}
\Bigr] . \notag
\end{split}
\end{equation}
As shown by \textcite{suess2015hierarchical,link2022non}, this hierarchy admits a pure-state unraveling of the form $\hat{\rho}_{\mathbf{m},\mathbf{n}}=\langle|\psi_{\mathbf{m}}\rangle\langle\psi_{\mathbf{n}}|\rangle_{\rm av}$ with
$\langle Z_t\rangle_{\rm av}=\langle Z_t Z_s\rangle_{\rm av}=0$ and $\langle Z_t Z_s^{*}\rangle_{\rm av}=C(t-s)$. Consequently, for $\psi_{\mathbf{n}}(t)\equiv\psi_{\mathbf{n}}[Z_t^{*};t]$ one obtains the hierarchy of pure states [HOPS;\textcite{suess14}]
\begin{align}
\partial_t \psi_{\mathbf{m}}(t)
&=\Bigl[-i\,\hat{H}_{s}+Z_t^{*}\hat{S}-\sum_{k=1}^{K} m_k\, z_k\Bigr]\psi_{\mathbf{m}}(t) \notag\\
&\quad+\sum_{k=1}^{K}\Bigl[m_k d_k\,\hat{S}\,\psi_{\mathbf{m}_k^{-}}(t)-\hat{S}\,\psi_{\mathbf{m}_k^{+}}(t)\Bigr] \,,
\end{align}
a nested set of stochastic Schrödinger equations in an extended state space. For factorized initial conditions, the HEOM-consistent boundary conditions $\psi_{\mathbf{m}}(0)=0$ for $\mathbf{m}\neq\mathbf{0}$ are automatically enforced, and the reduced density operator follows as $\hat{\rho}_s(t)=\langle |\psi_{\mathbf{0}}(t)\rangle\langle\psi_{\mathbf{0}}(t)|\rangle_{\rm av}$. Recently, \textcite{guo2025extended} proposed a generalized HOPS compatible with the decomposition in Eq.~\eqref{Eq:spectrum_dcp}, thereby providing an explicit unraveling of Eq.~\eqref{Eq:qdmess-density}.

By propagating pure states rather than density matrices, HOPS achieves a quadratic reduction in auxiliary-space dimensionality relative to approaches such as HEOM. The trade-off is the stochastic sampling: Ensemble averages must be converged over many noise realizations where the number of trajectories required for a chosen accuracy typically grows with propagation time, making long-time simulations costly. To accelerate convergence, particularly at strong system–bath coupling regimes, a nonlinear HOPS formulation based on the Girsanov-transformation has been developed \cite{suess14}. Recently, a displaced auxiliary-mode state has been proposed to mitigate the challenges of simulating highly excited bosonic environments \cite{muller2025quantum}. Combining HOPS with matrix product state techniques \cite{gao2022non,guo2024electronic,flannigan2022many} has further improved efficiency. 

\section{Thermofield Based Methods}
\label{Sec:thermofield}

The system--bath dynamics at finite temperature can also be studied using the thermofield-based wave function approach \cite{suzuki1991density,diosi1998non,yu2004non,borrelli2017simulation,gelin2021efficient,gelin2017thermal,vega2015thermofield}. There, the environmental Hilbert space is mirrored and then subjected to a thermofield Bogoliubov transformation \cite{umezawa1982thermo,takahashi1996thermo,henning1994diagonalization,umezawa1988micro}. This procedure transforms the environment’s initial mixed state into two virtual environments described by  vacuum states. Accordingly, the resulting thermofield Hamiltonian enables wavefunction-based methods to simulate finite-temperature environments including DMRG \cite{feiguin2005finite,vega2015thermofield,ren2022time,xu2023hierarchical,nuomin2022improving}, NRG \cite{schwarz2018nonequilibrium}, MCTDH \cite{fischer2021thermofield}, Davydov-Ansatz approaches \cite{zhao2023hierarchy}, NMQSD \cite{ritschel2015non}, and the TEDOPA \cite{vega2015thermofield}. 

The approach starts from system-bath Hamiltonian of the form
\begin{align}
    \hat{H}_0 = \hat{H}_s + \sum_{k=1}^K \Bigl[\omega_k \hat{b}_k^\dagger \hat{b}_k + g_k \hat{S} (\hat{b}_k^\dagger + \hat{b}_k) \Bigr] \,,
\end{align}
which is augmented by fictitious modes ('dark' modes) so that the total Hamiltonian reads $\check{H}_0 = \hat{H}_0 - \sum_k \omega_k \hat{c}_k^\dagger \hat{c}_k$. A set of collective modes is then obtained via a thermofield Bogoliubov transformation, i.e., 
\begin{equation}
\begin{aligned}
    \hat{a}_{1k} &= \sqrt{n_k + 1}\, \hat{b}_k - \sqrt{n_k}\, \hat{c}_k^\dagger \\
    \hat{a}_{2k} &= \sqrt{n_k + 1}\, \hat{c}_k - \sqrt{n_k}\, \hat{b}_k^\dagger
\end{aligned}
\end{equation}
with Bose-Einstein distribution \(n_k = 1/(e^{\beta\omega_k}-1)\). As a consequence, the mixed bath initial state is mapped onto a vacuum state (zero particle state) with respect to the operators \(\hat{a}_{1k}\) and \(\hat{a}_{2k}\) \cite{yu2004non,vega2015thermofield} and the corresponding coupling $g_{1k} = g_k \sqrt{n_k+1}$ and $g_{2k} = g_k\sqrt{n_k}$. The transformed Hamiltonian reads 
\begin{align}\label{Eq:thermo_hamilton}
    \tilde{H}_0 = & \hat{H}_s + \sum_{k=1}^K \Big[ \omega_k (\hat{a}_{1k}^\dagger \hat{a}_{1k} - \hat{a}_{2k}^\dagger \hat{a}_{2k}) \notag \\
    &+ g_{1k} \hat{S} (\hat{a}_{1k}^\dagger + \hat{a}_{1k}) + g_{2k} \hat{S} (\hat{a}_{2k}^\dagger + \hat{a}_{2k}) \Big] \, . 
\end{align}
It is this Hamiltonian that enables finite-temperature dynamics to be treated with \emph{pure-state wavefunction methods} rather than Liouvillian density-matrix equations.

However, for a continuum reservoir spectral density, the thermofield wave function method requires a significantly larger number of undamped modes to discretize it accurately, as detailed in \textcite{feiguin2005finite,vega2015thermofield,ren2022time,xu2023hierarchical,nuomin2022improving,schwarz2018nonequilibrium,fischer2021thermofield,zhao2023hierarchy,ritschel2015non}. This requirement poses  severe challenges for simulating the quantum dynamics over long-timescales. A benefit of the method is related to the fact that the system-bath entanglement entropy grows modestly with the number of effective modes in comparison to density matrix based methods, thereby allowing a lower bond dimension in matrix product state simulations [e.g., \textcite{ren2022time,le2024managing}].

The QD-MESS formulation is closely connected to the  thermofield methods, (cf.~Secs.~\ref{sec:qdmess}, \ref{sec:quasi-lindblad} and Appendix~\ref{sec:thermofield-transformation}), consistent with the well-known correspondence between thermofield dynamics and the Schwinger–Keldysh formalism  \cite{umezawa1982thermo,aurenche1992comparison,henning1994diagonalization,henning1995thermo,evans1992heisenberg,lundberg2021thermal,khanna2009thermal}. Namely, in the undamped limit \(\gamma_k = 0\), \eref{Eq:diagonal-basis} reduces to 
\begin{align}
  C(t)
  = \sum_{k=1}^{K} g_{k}^{2}\,
  \Big[(n_k{+}1)\,e^{-i\omega_k t} + n_k\,e^{+i\omega_k t}\Big]\,,
\end{align}
and the dynamics of \eref{Eq:Lindblad-tamascelli} reduces to a unitary one for a system plus pseudomodes. Therefore, it admits a pure-state unraveling reproducing the thermofield-based wave-function formulations. 

More generally, whenever the decomposition [\eref{Eq:spectrum_dcp}] implies that the dissipation matrix \(\boldsymbol{\mathcal{C}}\) in Eq.~\eqref{Eq:qdmess-Lindblad} vanishes, the bath correlation function simplifies to 
\begin{align}\label{Eq:dcp_nodamping}
    C(t) = \bm{\kappa}^\dagger e^{-i\bm{\Omega} t} \bm{\kappa} \,,
\end{align}
and the QD-MESS [Eqs.~\eqref{Eq:qdmess-sk} and \eqref{Eq:qdmess-Lindblad}] collapses to strictly unitary dynamics in system–mode space governed by the Hermitian embedded Hamiltonian \(\hat{H}_{0}\) in Eq.~\eqref{Eq:h-embedding}. The interaction topology of effective modes is encoded in the structure of \(\bm{\Omega}\), and particularly a tridiagonal $\bm{\Omega}$ corresponds to a chain topology, see next section.

As previously discussed, employing a damped mode representation for \(C(t)\) offers notable advantages in actual simulations though. However, the corresponding extension of the thermofield wave function approach necessitates its complete reformulation with close relations to stochastic unravelings, see \cite{ritschel2015non,suess2015hierarchical}.

\section{Chain Topology Methods}
\label{Sec:chain-topology}

As discussed in Sec.~\ref{subsec:discretize-bath}, chain-mapping methods employ an orthogonal transformation \cite{chin2010exact,bulla2005numerical} or numerical fitting \cite{medina2021few,monica2021accurate} to recast a system-bath interaction in ``star"-topology into a semi-infinite chain of harmonic modes with nearest-neighbor interactions. In this representation, the system couples exclusively to the first oscillator in the chain \cite{bulla03,bulla2005numerical,bulla2008numerical,chin2010exact,woods2014mappings,vega2015how,monica2021accurate,hughes2009effective,strasberg2016nonequilibrium}. In contrast to the original star geometry, the chain topology involves only short-range interactions, which typically lead to a slower growth of system-bath entanglement entropy \cite{lacroix2024connectivity,li2022fly,kohn2021efficient,nuomin2024efficient,schroder2019tensor}, however, see also \cite{wolf2014solving}. This feature facilitates efficient simulations using matrix product state techniques such as the DMRG and TEBD. Moreover, the chain representation may offer an intuitive picture about the flow of information and thermalization processes between system and  bath \cite{monica2021accurate,medina2021few,tamascelli2020excitation}.

In this setting, the cumulants of the bath autocorrelation function \(C(t)\) are intimately connected to the size of the so-called Krylov basis \cite{parker2019universal,viswanath1994recursion}. The principal bottleneck of naive chain-mapping methods is the growth of the chain length with simulation time \cite{jonsson2024chain,monica2021accurate,tamascelli2020excitation,gindensperger2007hierarchy,hughes2009effective,vega2015how,huang2025coupled} which in turn implies an exponential increase in the bond dimension of  matrix product states during the time evolution \cite{schollwock11}. These aspects render accurate long-time simulations, especially for driven systems, particularly challenging. To ensure the convergence of such methods, it is essential to estimate the asymptotic behavior of the Lanczos coefficients \cite{chin2010exact,woods2014mappings,parker2019universal} and to establish rigorous error bounds for state truncations \cite{trivedi2021convergence,trivedi2024a,jonsson2024chain,huang2024unified}.

Chain-mapping techniques have also been integrated within the framework of quantum master equations, both to address strong system-bath coupling \cite{garg1985effect,martinazzo2011communication,monica2021accurate,strasberg2016nonequilibrium,anto2023effective,iles2014environment,gindensperger2007hierarchy,hughes2009effective} and to extend simulation timescales \cite{monica2021accurate}. A well-known example is the one-dimensional reaction-coordinate mapping, which has been introduced on physical grounds already a while ago \cite{garg1985effect}. It incorporates a collective reservoir mode into an augmented system that couples weakly to a residual bath, thereby enabling a quantum master equation description.

\emph{Chain representation of bath correlation.---} 
The orthogonal polynomial method  \cite{chin2010exact,woods2014mappings} transforms a reservoir with spectral density $J(\omega)$ into a semi-infinite nearest-neighbor chain of effective modes. Finite-temperature effects can then be incorporated via a thermofield transformation \cite{yu2004non,vega2015thermofield}. More recently, an efficient chain representation of thermal baths is obtained directly from the noise spectrum $S_{\beta}(\omega)$ \cite{tamascelli2019efficient} by exploring a Bogoliubov transformation,  see also Sec.~\ref{Sec:thermofield}.

One defines the measure $d\mu(\omega) = S_\beta(\omega)d\omega/ 2\pi g_0^2$ with $g_0 = [C(0)]^{1/2}$ and, at zero-temperature, $S_\beta(\omega) = \theta(\omega) J(\omega)$. Then,  a set of orthonormal polynomials $\{\pi_n(\omega^l)\}$ ($n,l \in \mathbb{N}$) is introduced, i.e.\ $ \int_{-\infty}^\infty d\mu(\omega)\, \pi_m(\omega^l)\, \pi_n(\omega^l) = \delta_{mn}$, such that the polynomials obey the three-term recurrence relation 
\begin{equation}
   (\omega^l - \omega_n^l) \pi_n(\omega^l) = g_n\pi_{n-1}(\omega^l) + g_{n+1} \pi_{n+1}(\omega^l) \,,
\end{equation}
with initial conditions \(\pi_{-1}(\omega^l) = 0\) and \(\pi_0(\omega^l) = 1\). These parameters can be obtained numerically \cite{gautschi2005orthogonal,jonsson2024chain,viswanath1994recursion} or derived analytically in special cases \cite{chin2010exact,woods2014mappings}.
As a final step, one defines the Krylov basis functions $\mathcal{K}_n(t) = \int_{-\infty}^{\infty}d\mu(\omega)\, \pi_n(\omega^l)\, {e}^{-i\omega t}$, which satisfy the boundary condition $\mathcal{K}_n(0) = \delta_{n0}$. In this basis the bath noise spectrum gains a representation of the form 
\begin{align} \label{Eq:high-cdp}
    S_\beta(\omega) = \bm{\kappa}^\dagger \frac{i}{\omega^l\mathds{1} - \bm{\mathcal{E}} + i0^+} \bm{\eta} + \mathrm{H.c.} \,,
\end{align}
with effective parameters $\bm{\kappa} = \bm{\eta} = [g_0, 0, \ldots, 0]^T$ and  the semi-infinite tridiagonal matrix \(\bm{\mathcal{E}}\)  defined by
\begin{equation}\label{Eq:semi-inf-tri}
    \mathcal{E}_{m,n}
    \;=\;\omega_n^l\,\delta_{m,n}
    \;+\;g_n\,\delta_{m+1,n}
    \;+\;g_m\,\delta_{m,n+1} \,.
\end{equation}
In this manner, the bath correlation function is represented as a one-dimensional chain configuration of interacting modes in the constructed Krylov space such that what is considered as system couples exclusively to the first site of the chain. We mention that the QD-MESS describes a chain topology by choosing a proper representation of $\bm{\mathcal{E}}$ and the coupling vectors as we discuss in the sequel. 

\emph{Effective mode mapping.---}
In practical applications, the parameter \(l\) introduced above enables different chain realizations \cite{woods2014mappings}. For example, setting \(l=1\) in \eref{Eq:high-cdp} reduces it to the form of \eref{Eq:spectrum_dcp}. Consequently, the density-matrix QD-MESS [see Eqs.~\eqref{Eq:qdmess-sk} or \eqref{Eq:qdmess-Lindblad}] with a real-valued symmetric matrix \(\bm{\mathcal{E}}\) [see Eqs.~\eqref{Eq:semi-inf-tri} and \eqref{Eq:dcp_nodamping}] admits a pure-state unraveling, resulting in the thermalized time-evolving density operator with orthogonal polynomials [TEDOPA; \textcite{tamascelli2019efficient,lacroix2024mpsdynamics,chin2010exact,woods2014mappings,monica2021accurate}]. In this scenario, the chain Hamiltonian appears as 
\begin{align}\label{Eq:hparticle}
   \hat{H}_0 =&  \hat{H}_s + g_0\, \hat{S} (\hat{a}_0^\dagger + \hat{a}_0)  + \sum_{n=0}^{\infty} \Bigl[\omega_{n} \hat{a}_{n}^\dagger \hat{a}_{n} \notag \\
   & + g_{n+1}(\hat{a}_{n}^\dagger \hat{a}_{n+1} + \hat{a}_{n+1}^\dagger \hat{a}_{n})\Bigr] \,
\end{align}
describing mode interactions in the form of a tight-binding lattice with nearest-neighbor single-particle hopping. Recently, \textcite{lacroix2025making} employed the chain Hamiltonian to study its connection with quantum collision models.

Alternatively, choosing \(l=2\) in \eref{Eq:semi-inf-tri} leads to a chain of interacting modes when unraveling the influence functional in analogy to the procedure in Sec.~\ref{sec:qdmess}, in which we identify the  corresponding $\bm{G}(t)$ and its adjoint as the time- and anti-time-ordered Green's functions defined in coordinate space. The resulting Hamiltonian then reads
\begin{align}\label{Eq:hphonon}
    \hat{H}'_0 =& \hat{H}_s + g_0 \hat{S} (\hat{a}_0 + \hat{a}_0^\dagger) + \sum_{n=0}^{\infty}\Bigl[\omega_{n}\hat{a}_{n}^\dagger \hat{a}_{n} \notag \\
    &+ g_{n+1} (\hat{a}_{n}^\dagger + \hat{a}_{n})(\hat{a}_{n+1}^\dagger + \hat{a}_{n+1}) 
     \Bigr] \,,
\end{align}
with mode-mode interactions describing a coordinate-coordinate coupling. When setting $S_\beta(\omega) = J(\omega)\theta(\omega)$, this Hamiltonian aligns with the form given in \textcite{hughes2009effective,martinazzo2011communication,woods2014mappings,strasberg2016nonequilibrium}.

We emphasize that for these specific approaches the bath correlation function is determined solely by the Lanczos coefficients, rather than by any particular choice of the Krylov basis. In general, the Krylov space spanned by \(\mathcal{K}_n(t)\) is infinite, thus necessitating a truncation of the chain length to a finite number of \(K\) sites. However, such a truncation is prone to finite-size effects due to the scattering of excitations at the chain boundaries \cite{monica2021accurate,tamascelli2020excitation}, thereby limiting the accessible simulation time, particularly for systems subjected to non-periodic driving and strong reservoir retardation (non-Markovianity). Moreover, since chain-topology-based methods typically lack intrinsic damping, necessary chain lengths scale unfavorably  with simulation timescales \cite{vega2015how,monica2021accurate,mascherpa2020}. 

Several remedies have been proposed to overcome these drawbacks. For instance, a network of interacting damped modes has been introduced to arrive at a more efficient representation of reservoir spectral densities \cite{lednev2024lindblad,medina2021few,mascherpa2020}.  Alternatively, one may terminate the chain at a site $K$ and treat the residual bath via time-local master equations \cite{garg1985effect,martinazzo2011communication,monica2021accurate,strasberg2016nonequilibrium,anto2023effective,iles2014environment,hughes2009effective} as will be discussed next.

\emph{Non-Markov Chain Closure.---}
Once thermal bath modes are recast in a chain configuration, it appears to be suggestive to terminate the chain of length $K$ by declaring remaining modes as a residual bath. 
Its noise spectrum emerges out of a fractional  recursion relation~\cite{viswanath1994recursion,parker2019universal,marchesoni1983extension}, i.e.
\begin{equation}\label{eq:GF_recursion}
    {G}_{n,n}(\omega) = \frac{1}{\omega^l - \omega_n^l - g_{n+1}^2 \,{G}_{n+1,n+1}(\omega)},\; n \ge 0 \, ,
\end{equation}
where $G_{n,n}(\omega)$ denotes Green's function for a semi-infinite chain starting from site $n$. Thus, the chain comprising all sites smaller than a given $n$ together with the system forms an embedded system while the remainder of the chain, characterized by ${G}_{n,n}(\omega)$, $n>K$, acts as a thermal reservoir for site $K\equiv (n-1)$. More explicitly, the noise spectrum of this residual bath turns out to read (see Appendix~\ref{Sec:GreenFuncNoise} for details) 
\begin{subequations}
\begin{align}
    S_{\beta,n}(\omega) &= 
    - 2\,g_n^2\,\Im{G_{n,n}(\omega)} \,,\\[2mm]
    G_{n,n}(\omega) &= 
    \frac{1}{2\pi\,g_n^2}
    \int_{-\infty}^{\infty}\!d\omega'\,
    \frac{S_{\beta,n}(\omega')}{\omega - \omega' + i0^+}\,.
\end{align}    
\end{subequations}

By way of example, we focus on an embedded system composed of the original system plus a single mode: (i) For $l = 1$, choosing $S_{\beta,1}(\omega) = \gamma$ leads to an effective noise spectrum for the original system
\begin{equation}
    S_{\beta}(\omega) = 
    \frac{2\,g_0^2\,\gamma}{(\omega - \omega_0)^2 + \gamma^2} \, 
\end{equation}
as studied by \textcite{monica2021accurate,nazir2018the,correa2019pushing}. (ii) For $l = 2$, one chooses for the spectral density of the residual bath coupled to the chain mode
$J_1(\omega) = 2\gamma \omega {e}^{-\omega/\Lambda}\equiv S_{\beta\to \infty,1}(\omega)$ with sufficiently large $\Lambda$. 
In a fully reduced picture, i.e. seen from the system alone, this appears as the effective spectral density 
\begin{align}
    J(\omega) = \frac{4 g_0^2\gamma \omega}{(\omega^2 - \omega_0^2)^2 + 4\gamma^2\omega^2} \,.
\end{align}
This spectral density has been studied in the context of the reaction-coordinate chain mapping \cite{leggett1984quantum,garg1985effect,woods2014mappings,martinazzo2011communication,monica2021accurate,strasberg2016nonequilibrium,anto2023effective,iles2014environment,hughes2009effective}.

We have illustrated that, for the typical  one-dimensional chain configuration of the bath correlation function (with $\gamma = 0$), the QD-MESS formalism simplifies to a Schr\"odinger wave-function–like method, which remains valid even at finite temperatures. Apart from the star and the chain topology, see Fig.~\ref{fig:mapping_model}, the QD-MESS framework further enables hybrid topologies that combine features of these limiting cases via a structured matrix \(\bm{\mathcal{E}}\) and coupling vectors \(\bm{\kappa}\) and \(\bm{\eta}\), cf. \textcite{huh2014linear,tamura2007nonadiabatic,cederbaum2005short,gindensperger2007hierarchy,nazir2018the,nuomin2024efficient}. However, it remains an open question which topology permits a minimal effective number of bath modes while simultaneously inducing a sufficiently slow growth of system–bath entanglement entropy during the time evolution \cite{lacroix2024connectivity,li2022fly,kohn2021efficient,nuomin2024efficient,schroder2019tensor}. Closely related to these issues are the challenges of performing controlled truncations at manageable chain lengths without sacrificing accuracy and of reconstructing Green's functions from truncated (incomplete) continued fractions \cite{viswanath1994recursion}.

\section{Non-Markovian Perturbative Treatments}
\label{Sec:MasterEquation}

The QD-MESS platform offers also a convenient starting point for  systematic weak coupling perturbative treatments while keeping non-Markovian features. This development is in line with recent activities to formulate quantum master equations for non-Markovian dynamics, see e.g.\  \textcite{breuer16,vega17,brian2021generalized,mulvihill2021a,gonzalez2024tutorial}.  

One starts with \eref{Eq:qdmess-density} by defining  bilinear Liouville space superoperators  $\bm{X}_{c} \hat\rho = \{\bm{\kappa}^\dagger \hat{\bm{a}} + \hat{\bm{a}}^\dagger\bm{\eta}, \hat{\rho} \}_{\star}$ and $\bm{X}_{q} \hat\rho = [\hat{\bm{a}}^\dagger\bm{\eta}, \hat{\rho}]_{\star}$, so that in the interaction picture \eref{Eq:qdmess-density} turns into
\begin{align}\label{Eq:formal-liouville}
\dot{\hat\rho}(t) &= -i\,\mathcal{L}_{\rm sm}(t)\,\hat\rho(t) \notag \\
&= -i\bigl[\mathcal{S}_{q}(t)\,X_{c}(t) + \mathcal{S}_{c}(t)\,X_{q}(t)\bigr] \hat\rho(t)\,.
\end{align}
The formal solution of this equation is given by $\hat\rho(t)=\mathcal{G}(t,t_0)\,\hat\rho(t_0)$ with the propagator \(\mathcal{G}(t,t_0)\) obeying the Dyson equation
\begin{equation}\label{Eq:dyson-relation}
\mathcal{G}(t,t_0)=\mathcal{G}_0(t,t_0)-i\int_{t_0}^t d\tau\,\mathcal{G}_0(t,\tau)\,\mathcal{L}_{\rm sm}(\tau)\,\mathcal{G}(\tau,t_0)\,, \notag
\end{equation}
which includes the bare system propagator \(\mathcal{G}_0(t,\tau)\). Upon iteratively solving the Dyson equation one incorporates systematically  higher order correlations in the system-bath coupling.

Let us consider the second order approximant to \(\mathcal{L}_{\rm sm}\).  After tracing out the effective modes one obtains the generalized second-order master equation
\begin{multline}\label{Eq:redfieldplus}
\dot{\hat\rho}_s(t)=-\int_{0}^{t}d\tau\Bigl\{\Bigl[\hat{S}(t),\,C(t-\tau)\, \hat{S}(\tau)\,\hat\rho_s(\tau)\Bigr]\\
-\Bigl[\hat{S}(t),\,C^\ast(t-\tau)\,\hat\rho_s(\tau)\, \hat{S}(\tau)\Bigr]\Bigr\}\,.
\end{multline}
Unlike the standard Redfield equation, \eref{Eq:redfieldplus} retains the explicit time dependence of the density operator \(\hat\rho_s(\tau)\), thereby capturing memory effects inherent in non-Markovian dynamics. It can be transformed to Lindblad form \cite{gulacsi2023signatures} exhibiting negative decay rates. Within the generalized HEOM \eref{Eq:baryheom4}, this second order evolution appears naturally when 
setting all ADOs with $\sum_k (n_k+m_k)>1$ to zero \cite{xu2022minimal,xu2021heat}. 

To go beyond second order is straightforward in principle but rather tedious in practice \cite{crowder2024invalidation,magazzu2024unified,wu2013higher,jang2002fourth,xu2018convergence,dan2022generalized}. In addition, the emerging multi-dimensional integrals
are often numerically challenging and prone to errors. To address these issues,  diagrammatic techniques have been employed \cite{cohen2015taming,chen2017inchworm,gull2011continuous,gu2024diagrammatic}. An alternative route is based on the HEOM representation and truncates the ADOs such that contributions with \(\sum_k (n_k+m_k) > L\) are neglected.  The truncated HEOM  becomes then equivalent to a \(2L\)th-order time-nonlocal generalized quantum master equation \cite{xu05,schroder2007reduced}, thus enabling both efficient and accurate computations.

\begin{table*}[t]
\caption{Key features of selected non-perturbative approaches for simulating open quantum systems and their relation the QD-MESS. Each approach (i) specifies a system–mode topology within an extended state space, (ii) uses a specific representation of the reservoir modes, and (iii) extracts the reduced density operator after propagation from a respective initial state. }
\label{tab:method-features}
\renewcommand{\arraystretch}{1.2}
\setlength{\tabcolsep}{4pt}
\begin{tabularx}{\textwidth}{
l
>{\centering\arraybackslash}p{4cm}
>{\raggedright\arraybackslash}X
>{\raggedright\arraybackslash}p{2.6cm}
>{\raggedright\arraybackslash}p{4.0cm}} %
\hline\hline
Method & Mode topology / Effective reservoir coupling  & Reservoir representation  & Reduction to $\hat{\rho}_{s}$ & Link to QD-MESS \\
\hline
HEOM & Network / Complex &
Fock state &
Vacuum projection &
First major form in number basis \\[5pt]
Lindblad--pseudomode & Network / Positive &
Density operator &
Trace of mode space &
Specific case of second major form\\[5pt]
Brownian motion & -- / Positive &
Phase space &
$\int\! dxdp\,\hat{\rho}(x,p)$ &
Weyl correspondence \\[5pt]
SLN & -- / -- &
Probability space &
$\langle \hat{\rho}_{z}\rangle_{\rm av}$
&
Ornstein-Uhlenbeck decomposition \\[5pt]
HOPS & Network / Complex &
Probability $+$ Fock &
$\langle|\psi_{\mathbf{0}}\rangle\langle\psi_{\mathbf{0}}|\rangle_{\rm av}$ &
Stochastic unraveling of first major form \\[5pt]
TEDOPA & Chain / Positive &
Matrix product state &
Trace of mode space &
Specific case of second major form \\
\hline\hline
\end{tabularx}
\end{table*}
%

\section{Summary and Outlook}
\label{sec:summary}

Over the past decades, significant progress has been achieved in simulating open quantum systems, particularly concerning non-Markovian system-bath dynamics. These advances have been driven by the necessity to follow experimental progress in monitoring and manipulating  complex quantum systems by external fields with growing accuracy. Accordingly, approaches have been put forward which reflect the challenges of a broad range of specific fields such as  
 solid-state and condensed matter physics, quantum optics, atomic and molecular physics, chemical physics, and, more recently, quantum information science. 

 Recent advances provide convincing evidence that embedding methods (also termed hybrid methods) offer considerable advantages in terms of broad applicability, numerical stability, and computational efficiency in comparison to approaches working in reduced system Hilbert space or those working in full Hilbert of system and reservoir. The key idea of embedding methods is to construct an extended state space which includes the system of interest and effective reservoir modes as practical means to reconstruct environmental noise and dynamical response in combination with a proper propagation scheme for the density matrix. This can be done from two perspectives, namely, either via the introduction of auxiliary states as a mere means of computational convenience to effectively represent the reservoir or via the construction of a modestly sized reservoir model based on desired physical analogies/properties. 

 In this review, we present a unified framework for quantum dissipation within a minimally extended state space (QD-MESS) which allows us to reveal the relationships between several prominent methodologies that have been formulated based on either of the two perspectives. Our intention has been to provide the reader with a concise yet transparent overview of the major recent developments guided by a `holistic' approach. We demonstrate the common ground and features of these methods and at the same time their differences when it comes to details and numerical performances. Within the scope of this Colloquium, we decided to focus on major approaches such as the hierarchical equations of motion, the Lindblad-pseudomode formulation, quantum Brownian motion master equations, stochastic and chain mapping approaches. Their key distinguishing features are summarized in Table.~\ref{tab:method-features}.

Although these classes of methods are, as we show, intimately related and even formally equivalent under suitable transformations, their computational requirements and simulation efficiencies vary significantly depending on the specific parameter regimes. Therefore, there is no `gold standard' but the optimal simulation has to be chosen according to the problem under consideration. We hope that this work will support the reader in identifying the appropriate method and trigger new developments. 

This Colloquium focuses on embedding methods but we want to briefly mention alternative strategies and refer the reader to the existing literature for further details. In order to overcome the `curse of dimensionality' in simulating many-body quantum systems in full Hilbert space, techniques based on tensor network and neural network representations have emerged in the last years. For systems governed by time-local equations, methods such as the Numerical Renormalization Group (NRG), Time-Evolving Block Decimiation (TEBD), and Density Matrix Renormalization Group (DMRG) enable simulations for system-reservoir interaction in chain-topology. For Hamiltonians exhibiting long-range interactions, the Time-dependent Variational Principle [TDVP; \textcite{lubich15,haegeman2011time,shi2018efficient}] and the multi-layer MCTDH algorithm offer distinct advantages. Alternatively, quantum dynamics in the path integral representation can be calculated using the Time-evolving Matrix Product Operator [TEMPO; \textcite{strathearn2018efficient}] and the Tensor Network Path Integral [TNPI; \textcite{bose2022multisite}] approaches. Recently, the neural network quantum state representations have been proposed for Markovian dissipation \cite{norambuena2024physics,dugan2023qflow,luo2022autoregressive,nagy2019variational,hartmann2019neural,vicentini2019variational,yoshioka2019constructing}, and extended to the non-Markovian regime \cite{cao2024neural}. 
The strategy of embedding methods is closely related to the context of environment engineering \cite{feist2020macroscopic}. Dynamical Mean-Field Theory [DMFT; \cite{aoki2014nonequilibrium}] is also based on a somewhat similar strategy for simulating strongly correlated systems by mapping the problem onto a quantum impurity model embedded in a self-consistently determined effective Gaussian bath \cite{arrigoni2013nonequilibrium,hou2014hierarchical,scarlatella2021dynamical,bertrand2025turning}. 

Where do we go? Where are the challenges?
The need for high-precision predictions for the dynamics of tailored quantum systems and devices is still pressing given also technological developments. For example, quantum information processing relies on the precise control of quantum states even where the presence of environmental noise is inevitable. While delicate quantum properties are inherently susceptible to such noise, dissipation engineering techniques can be harnessed as essential tools for quantum measurement, state preparation, state stabilization, quantum error mitigation, and quantum simulation.
Further, the advent of quantum state manipulation, quantum thermodynamics, and the investigation of open quantum few to many-body systems come with formidable challenges that originate from the interplay of time-dependent driving, long-range correlations, also  mediated by environments, multiple nonlinear degrees of freedom, pronounced non-Markovian effects, and strong system--bath hybridization. Addressing these issues requires even more efficient non-perturbative simulation tools. To formulate them, fundamental theory has to go hand in hand with Hamiltonian modeling and numerical algorithms.

\begin{acknowledgments}
JA and JTS are very grateful to H. Grabert and U. Weiss for numerous discussions.
We gratefully acknowledge financial support from the IQST and the German Research Foundation (DFG) under AN336/12-1 (FOR2724) and AN336/17-1, from the BMBF within the project QSolid, and from the State of Baden-W\"urttemberg through the network KQCBW. V.V. acknowledges support from the Academy of Finland Centre of Excellence program (project No. 336810) and THEPOW (project No. 349594), and from the European Research Council under Advanced Grant No.~101053801 (ConceptQ).
\end{acknowledgments}

\appendix

\section{Green's Function Representation of Noise Spectra}
\label{Sec:GreenFuncNoise}

Here we collect some basic information about Green's functions and their relation to the spectral noise power used in Secs.~\ref{Sec: OpenQuantumSystems}, \ref{sec:qdmess}, and \ref{Sec:chain-topology}.

In this Review, we define the Fourier transform
\begin{align}
    F(\omega) &= \int_{-\infty}^{\infty} dt\, f(t)\, {e}^{i\omega t}\,,\\[1mm]
    f(t) &= \frac{1}{2\pi}\int_{-\infty}^{\infty} d\omega\, F(\omega)\, {e}^{-i\omega t}\,.
\end{align}
Thus, the correlation function \(C(t)\) and the noise spectrum \(S_\beta(\omega)\) form a Fourier pair in the main text.

We assume that the operator space spanned by single-particle Green's functions is well defined and that the two-point time correlation function for a stationary Gaussian bath obeys $C(\tau,\tau') = C^\ast(\tau',\tau)$. In  Green's function representation, the bath correlation function can be constructed as specified in Eq.~\ref{Eq:corr_dcp} so that its Fourier transform reads as in Eq.~\ref{Eq:spectrum_dcp}.

Green's function may also be expressed in its spectral representation,
\begin{align}
\bm{\kappa}^\dagger\,\bm{G}(\omega)\,\bm{\eta} = \frac{1}{2\pi}\int_{-\infty}^{\infty} d\omega'\,\frac{S_\beta(\omega')}{\omega-\omega'+i0^+}\,,
\end{align}
so that its Fourier transform satisfies
\begin{align}
    \frac{1}{2\pi}\int_{-\infty}^{\infty} d\omega\,\bm{\kappa}^\dagger\,\bm{G}(\omega)\,\bm{\eta}\, {e}^{-i\omega t} = -i\,\theta(t)\,C(t)\,.
\end{align}
These relations have been used iteratively in Sec.~\ref{Sec:chain-topology} to perform the non-Markovian chain closure.

\section{Path-Integral derivation of QD-MESS}
\label{Sec:skpi}
We here present an alternative derivation of the first major form of the QD-MESS, Eq.~\eqref{Eq:qdmess-density} in Sec.~\ref{sec:qdmess},  using the path integral formalism.

The starting point is the path integral representation to formulate Gaussian quantum dissipation non-perturbatively as pioneered by Feynman and Vernon \cite{feynman63} and further extended in a large body of work \cite{weiss12,grabert1988quantum,leggett87}. The advantage of this approach is that the reservoir degrees of freedom can be integrated out exactly and operator ordering problems do not appear explicitly.

\emph{Path integral representation.---}
We consider the total Hamiltonian \eref{Eq:htot}
and focus on an initially factorized state $\hat{\rho}_{\rm sb}(0) = \hat\rho_s(0)\otimes \hat\rho_{b}(0)$. After eliminating the Gaussian integrals corresponding to the reservoir degrees of freedom the reduced density operator $\hat{\rho}_s(t)$ attains a compact representation as a functional integral over paths along a forward-backward contour in time, i.e.\
\begin{equation}\label{Eq:path-integral}
  \rho_s^\pm(t)=\int\!\mathcal{D}[s_+,s_-] \, \mathcal{A}_s[s_+,s_-] \, e^{-i\Phi[s_+,s_-]}\, \rho_s^\pm(0) \, .
\end{equation}
Here, the bare system part is denoted by $\mathcal{A}_s[s_+,s_-]$  
and $s_\pm(t)$ are real-valued paths on the forward ($+$) and backward ($-$) branches, respectively, with boundary conditions indicated through the matrix elements $\rho_s^{\pm} = \langle s_+ |\hat{\rho}_s|s_-\rangle$. The impact of the environment is fully captured by the so-called Feynman-Vernon influence phase \cite{feynman63,weiss12}  
\begin{align}\label{Eq:influence-phase}
    \Phi[s_+,s_-] &= -i\int_0^t\!\! d\tau \!\!\int_0^\tau\!\! du\, \left[s_+(\tau)-s_-(\tau)\right] \notag \\
    &\times \left[C(\tau\!-\!u)s_+(u) \!-\! C^\ast(\tau\!-\!u)s_-(u)\right] \,. 
\end{align}
For convenience, one now introduces classical/quantum paths $s_{c/q} =(s_+ \pm s_-)/\sqrt{2}$ so that one arrives at  
\begin{equation}\label{Eq:IFsuperop_pi}
    \Phi[s_+,s_-] \!=\! \int_0^t\!\! d\tau du 
    \begin{bmatrix}
        s_{q}(\tau) \!&\! s_{c}(\tau) 
    \end{bmatrix} 
\bm{\Sigma}(\tau\!-\!u)
    \begin{bmatrix}
        s_{q}(u) \\ s_{c}(u)
    \end{bmatrix} ,
\end{equation}
where the self-energy kernel $\bm{\Sigma}(\tau\!-\!u)$ is identical to the one defined in \eref{Eq:self-energy_matrix}. 

\emph{Unraveling influence functional.---}
Given the representation of the spectral noise power in terms of Green's functions, see Sec.~\ref{Sec:gfrbs}, the goal is to properly unravel the influence functional in order to cast the time-nonlocal path integral expression into an equivalent time-local equation of motion in an extended Liouville space. For this purpose, substituting \eref{Eq:corr_dcp} into \eref{Eq:self-energy_matrix} yields a left-right factorization where the time dependence is contained in a block-diagonal term
\begin{align}\label{Eq:selfEnergy_factorization}
    \bm{\Sigma}(t) &\!=\! \begin{bmatrix}
        \bm{\kappa}^\dagger & \bm{\eta}^\dagger \\  \bm{0} & -\bm{\eta}^\dagger
    \end{bmatrix}
    \begin{bmatrix}
        \bm{G}(t) & \bm{0} \\ \bm{0} & -\bm{G}^\dagger(-t)
    \end{bmatrix}
    \begin{bmatrix}
        \bm{\eta} & \bm{\eta} \\ \bm{\kappa} & \bm{0}
    \end{bmatrix} \notag \\
    &\! \equiv  \bm{U}^\dagger\, \mathbb{G}(t)\, \bm{V} \,.
\end{align}
This way, the influence functional takes the form $e^{-i\Phi(t)} = \exp\{\!-i\! \int_0^t d\tau du\, \bm{s}_u^\dagger(\tau)\, \mathbb{G}(\tau\!-\!u)\,  \bm{s}_v(u) \}$, where the $2K\times 2$ matrices $\bm{U}$ and $\bm{V}$ have been absorbed into the sources by defining the row and column vectors $\bm{s}_u^\dagger(t) = [s_q(t)\;s_c(t)]\,\bm{U}^\dagger$ and $\bm{s}_v(t) = \bm{V}\,[s_q(t)\,s_c(t)]^T$. 
The time non-locality in the influence functional can then be unraveled by means of the Gaussian identity \cite{altland2010condensed,sieberer2016keldysh} 
\begin{align}\label{Eq:GaussianIdentity}
&\int \mathcal{D}[\bm{\phi}^\dagger,\bm{\phi}]\, e^{i\int_0^t d\tau 
\left[ \bm{\phi}^\dagger(\tau)\mathbb{K}(i\partial_\tau)\bm{\phi}(\tau) -\bm{\phi}^\dagger(\tau) \bm{s}_{v}(\tau) - \bm{s}_{u}^\dagger(\tau) \bm{\phi}(\tau)
\right] } \notag \\
&=e^{-i\int_0^t d\tau du\,\bm{s}_{u}^\dagger(\tau)\,\mathbb{G}(\tau-u)\,\bm{s}_{v}(u)} \,,
\end{align}
where the normalization factor is absorbed in the measure \(\mathcal{D}[\bm{\phi}^\dagger,\bm{\phi}]\) and $\mathbb{K}(i\partial_\tau) \mathbb{G}(\tau) = \delta(\tau)\mathds{1}$ with
\begin{align}
    \mathbb{K}(i\partial_t) = \begin{bmatrix}
         i\partial_t - \bm{\mathcal{E}} & \bm{0} \\
         \bm{0}  & -i\partial_t + \bm{\mathcal{E}}^\dagger
    \end{bmatrix} \,.
\end{align}
The term in \eref{Eq:GaussianIdentity} containing $\mathbb{K}(i\partial_\tau)$ can be identified with the Lagrangian of the coherent-state path integral of $2K$ free harmonic modes \cite{negel1998quantum}.

By inserting \ref{Eq:GaussianIdentity} into \ref{Eq:path-integral} and introducing the vectorial notation $\bm{\phi}(t)=[\bm{\phi}_+(t)~ \bm{\phi}_-(t)]^T$ with $\bm{\phi}_{\pm} = [\phi_{1,\pm}\;\ldots\;\phi_{K,\pm}]^T$, the reduced density operator is represented in extended space as
\begin{equation}
\rho_s^\pm(t) =\! \int\!\mathcal{D}[\bm{\phi}^\dagger, \bm{\phi}]\, \mathcal{D}[s_{c}, s_{q}] w[\bm{\phi}^\dagger, \bm{\phi}] \mathcal{A}[s_{c},s_{q};\bm{\phi}^\dagger,\bm{\phi}]\, \rho_s^\pm(0) \end{equation}
with
\begin{align}
\mathcal{A} = \mathcal{A}_s \exp\Big\{-i\!\int_0^t\! d\tau\,
\mathscr{L}_{\rm sm}(\bm{\phi}_+,\bm{\phi}_+^\dagger;\bm{\phi}_-,\bm{\phi}_-^\dagger) \Big\}\,,
\end{align}
the system-mode interaction Lagrangian 
\begin{align}
    \mathscr{L}_{\rm sm} =&\, \bm{\phi}^\dagger(\tau) \bm{s}_v(\tau) + \bm{s}_u^\dagger(\tau) \bm{\phi}(\tau) \notag \\
    = &\, s_q(\tau) \bm{\kappa}^\dagger {\bm{\phi}}_+(\tau)
+ \sqrt{2} s_+(\tau) {\bm{\phi}}_+^\dagger(\tau) \bm{\eta} \notag\\
& + s_q(\tau) {\bm{\phi}}_-^\dagger(\tau) \bm{\kappa}
- \sqrt{2} s_-(\tau) \bm{\eta}^\dagger {\bm{\phi}}_-(\tau)\, ,
\end{align}
and the effective mode path measure
\begin{align}\label{Eq:aupi-weights}
    w[\bm{\phi}^\dagger, \bm{\phi}] = {e}^{i\!\int_0^t\! d\tau [\bm{\phi}^\dagger(\tau)\mathbb{K}(i\partial_\tau)\bm{\phi}(\tau)]}.
\end{align}
The latter is the ``free-field'' action of bosonic field excitations from the vacuum, with Green's functions $\bm{G}(t)$ and $\bm{G}^\dagger(-t)$, restricted to $K$ modes, and with two independent copies for ``$+$'' and ``$-$'' paths. The boundary conditions of the coherent-state path integrals correspond to {\em vacuum-state boundary conditions}%
\footnote{This results from a careful consideration of the continuum limit of time-discrete path integrals.}. 
Assigning raising and lowering operators $\hat{a}_{k,\pm}^\dagger$ and $\hat{a}_{k,\pm}$ to the pure-state modes described by the coherent-state paths $\phi_{k,\pm}^\ast$ and $\phi_{k,\pm}$ that acts on the density operator from left and right, the extended system-mode Liouville space dynamics is now equivalent to the first major form of QD-MESS, \eref{Eq:qdmess-density}, in the main text.

\section{Lindblad-pseudomode transformation}
\label{sec:thermofield-transformation}
Here we implement Bogoliubov transformations to relate the various Lindblad–pseudomode equations discussed in Sec.~\ref{sec:quasi-lindblad}. This subsection also  underpins the discussions in Secs.~\ref{sec:quasi-lindblad}, \ref{Sec:thermofield}, and \ref{Sec:chain-topology}.

For notational simplicity (without loss of generality), we illustrate the case \(K=1\) following \eref{Eq:diagonal-basis}:
\begin{align} \label{Eq:single-ct-app}
    C(t\geq 0) &= g^2\Bigl[(n_\beta+1)\,e^{-i\omega t} + n_\beta\,e^{i\omega t}\Bigr]e^{-\gamma t} \notag\\[1mm]
    &= i\,\bm{\kappa}^\dagger\,\bm{G}^R(t)\,\bm{\eta}\,,
\end{align}
where the two exponential terms are treated as distinct modes. In particular, we choose $\bm{\kappa}^\dagger = \bm{\eta} = [g_1~g_2]$ with $g_1 = g(n_\beta + 1)^{1/2}$ and $g_2 = g (n_\beta)^{1/2}$, and matrix \[\bm{\mathcal{E}} = \begin{bmatrix}
    \omega - i\gamma & 0 \\ 0 & -\omega-i\gamma
\end{bmatrix}.\]  By these settings, Eqs.~\eqref{Eq:qdmess-sk} or \eqref{Eq:Lindblad-tamascelli} yield the Lindblad--pseudomode master equation
\begin{align}\label{eq:master-original}
    \dot{\hat{\rho}}(t) &= -i\,[\hat{\tilde{H}}_0,\,\hat{\rho}(t)] + \gamma\Bigl(2\hat{a}_1\,\hat{\rho}(t)\,\hat{a}_1^\dagger - \{\hat{a}_1^\dagger \hat{a}_1,\,\hat{\rho}(t)\}\Bigr) \nonumber\\[1mm]
    &\quad + \gamma\Bigl(2\hat{a}_2\,\hat{\rho}(t)\,\hat{a}_2^\dagger - \{\hat{a}_2^\dagger \hat{a}_2,\,\hat{\rho}(t)\}\Bigr)\,,
\end{align}
with the embedded system Hamiltonian
\begin{align}\label{eq:hamiltonian-original}
    \hat{\tilde{H}}_0 &= \hat{H}_s + \omega\,(\hat{a}_1^\dagger \hat{a}_1 - \hat{a}_2^\dagger \hat{a}_2) \nonumber\\[1mm]
    &\quad + g_1\,\hat{S}\,(\hat{a}_1^\dagger + \hat{a}_1) + g_2\,\hat{S}\,(\hat{a}_2^\dagger + \hat{a}_2)\,.
\end{align}
In simulations, the effective modes are initialized in the vacuum state and the reduced density operator is obtained as $\hat{\rho}_s(t)={\rm Tr}_{\rm b}\{\hat{\rho}(t)\}$. 

\emph{Thermofield Bogoliubov Transformation.---}
Implementing a two-mode squeezing transformation,
\begin{equation}\label{eq:bog-trans}
\begin{aligned}
    \hat{a}_1 &= \sqrt{n_\beta+1}\,\hat{b} - \sqrt{n_\beta}\,\hat{c}^\dagger\,,\\[1mm]
    \hat{a}_2 &= \sqrt{n_\beta+1}\,\hat{c} - \sqrt{n_\beta}\,\hat{b}^\dagger\,,
\end{aligned}
\end{equation}
yields the transformed Hamiltonian $\hat{\check{H}}_0 = \hat{H}_s + \omega\,\hat{b}^\dagger \hat{b} + g\,s\,(\hat{b}+\hat{b}^\dagger) - \omega\,\hat{c}^\dagger \hat{c}$. In the squeezed frame, the Lindblad dissipator is given by
\begin{align}\label{eq:dissipator}
    \mathcal{D}\hat{\rho} &= \sum_{j,k=1}^4 \mathcal{C}_{jk}\Bigl(2\hat{F}_j\,\hat{\rho}\,\hat{F}_k^\dagger - \{\hat{F}_k^\dagger \hat{F}_j,\,\hat{\rho}\}\Bigr)\,,
\end{align}
where the set of operators is $(\hat{F}_1,\hat{F}_2,\hat{F}_3,\hat{F}_4) = (\hat{b},\; \hat{c},\; \hat{b}^\dagger,\; \hat{c}^\dagger)$ and the Kossakowski matrix is
\begin{equation}\label{eq:kossakowski}
\mathcal{C}=\gamma
\begin{bmatrix}
\cosh^2\theta & 0 & 0 & \frac{1}{2}\sinh2\theta\\[1mm]
0 & \cosh^2\theta & \frac{1}{2}\sinh2\theta & 0\\[1mm]
0 & \frac{1}{2}\sinh2\theta & \sinh^2\theta & 0\\[1mm]
\frac{1}{2}\sinh2\theta & 0 & 0 & \sinh^2\theta
\end{bmatrix}\,.
\end{equation}
Tracing out the \(c\)-mode produces an effective Lindblad equation for the \(b\)-mode:
\begin{align}\label{eq:master-transformed}
    \dot{\hat{\rho}}(t) &= -i\,[\hat{H}_0,\,\hat{\rho}(t)] + \gamma\,(n_\beta+1)\Bigl(2\hat{b}\,\hat{\rho}(t)\,\hat{b}^\dagger - \{\hat{b}^\dagger \hat{b},\,\hat{\rho}(t)\}\Bigr) \nonumber\\[1mm]
    &\quad + \gamma\,n_\beta\Bigl(2\hat{b}^\dagger\,\hat{\rho}(t)\,\hat{b} - \{\hat{b}\,\hat{b}^\dagger,\,\hat{\rho}(t)\}\Bigr)\,,
\end{align}
with the effective Hamiltonian
\begin{equation}\label{Eq:hsb-app}
    \hat{H}_0 = \hat{H}_s + \omega\,\hat{b}^\dagger \hat{b} + g\,\hat{S}\,(\hat{b}+\hat{b}^\dagger)\,.
\end{equation}
The relations between Eqs.~\eqref{Eq:Lindblad-tamascelli} and \eqref{eq:lindblad-transformed} become evident, as stated in the main text. In simulations, the state of the \(b\)-mode is initialized according to \(n_\beta\), and the reduced density operator is given by $\hat{\rho}_s(t)={\rm Tr}_{\rm b}\{\hat{\rho}(t)\}$. 

We note that for finite-temperature open quantum systems, and regardless of damping effects, one typically applies the thermofield transformation to Hamiltonian \eref{Eq:hsb-app} to obtain Hamiltonian \eref{eq:hamiltonian-original} \cite{yu2004non,borrelli2017simulation,gelin2017thermal,vega2015thermofield}. In this latter case, the thermal equilibrium state acts as a thermal vacuum state, thereby enabling simulation of the system–bath dynamics via the Schr\"odinger equation, as described in Sec.~\ref{Sec:thermofield}.

\bibliographystyle{apsrmp4-1}
\bibliography{quantum}

\end{document}